\documentclass[trackchanges]{aastex7}
\usepackage{CJK}
\usepackage{multirow}
\usepackage{booktabs} 
\usepackage{verbatim}

\usepackage{amsmath}

\begin{document}
\begin{CJK*}{UTF8}{gbsn}
\title{Bayesian Geometrical Modeling of IXPE Polarization Angle Curves of the Magnetars 1E 2259+586 and 1E 1547.0$-$5408}

\author[0000-0003-0325-6426]{B. P. Li (李彪鹏)}
\affiliation{State Key Laboratory of Radio Astronomy and Technology, Xinjiang Astronomical Observatory, Chinese Academy of Sciences, 150, Science-1 Street, Urumqi, Xinjiang, 830011, China}
\affiliation{University of Chinese Academy of Sciences, No. 19 Yuquan Road, Beijing 100049, China}
\email{libiaopeng@xao.ac.cn}

\author[0000-0002-0138-3360]{Z. F. Gao (高志福)}
\affiliation{State Key Laboratory of Radio Astronomy and Technology, Xinjiang Astronomical Observatory, Chinese Academy of Sciences, 150, Science-1 Street, Urumqi, Xinjiang, 830011, China}
\email[show]{zhifugao@xao.ac.cn}

\author[]{W. Q. Ma (马文琦)}
\affiliation{State Key Laboratory of Radio Astronomy and Technology, Xinjiang Astronomical Observatory, Chinese Academy of Sciences, 150, Science-1 Street, Urumqi, Xinjiang, 830011, China}
\affiliation{University of Chinese Academy of Sciences, No. 19 Yuquan Road, Beijing 100049, China}
\email{mawenqi@xao.ac.cn}

\author[]{W. F. Zhang (张伟丰)}
\affiliation{State Key Laboratory of Radio Astronomy and Technology, Xinjiang Astronomical Observatory, Chinese Academy of Sciences, 150, Science-1 Street, Urumqi, Xinjiang, 830011, China}
\affiliation{University of Chinese Academy of Sciences, No. 19 Yuquan Road, Beijing 100049, China}
\email{zhangweifeng@xao.ac.cn}

\author[]{L. C. Garcia de Andrade}
\affiliation{Cosmology and Gravitation Group, Departamento de F\'isica Te\'orica-IF-UERJ, Rua S\~ao Francisco Xavier 524, Maracan\~a, Rio de Janeiro, RJ CEP:20550, Brazil}
\affiliation{Institute for Cosmology and Philosophy of Nature, Kri\v{z}evci, Croatia}
\email{luizandra795@gmail.com}

\begin{abstract}
X-ray polarimetry directly probes the radiation geometry and large-scale magnetic configuration of magnetars. We present a uniform Bayesian comparison between a dipole-dominated classical rotating vector model (CRVM) and a modified rotating vector model (MRVM) including a first-order magnetospheric twist correction. The models are applied to the phase-resolved IXPE polarization position angle (PA) curves of 1E~2259+586 and 1E~1547.0$-$5408. Parameters are inferred with a PA-level likelihood, and the models are compared using $\chi^2$, AIC, BIC, and Bayesian evidence. For 1E~1547.0$-$5408, we also test radio-derived geometrical constraints using radio-informed priors and radio-fixed fits. The current IXPE PA data for both sources are consistent with a dipole-dominated geometry and do not require a significant global twist. The MRVM gives only a marginal improvement for 1E~2259+586, with a Bayes factor of $\simeq3.3$, and no meaningful improvement for 1E~1547.0$-$5408, with a Bayes factor of $\simeq1.29$. We confirm that for 1E 1547, the nearly aligned radio geometry is not ruled out, but the radio RVM central geometry is not preferred by the X-ray PA data alone. The two sources show different impact angles, suggesting that magnetar X-ray polarization diversity reflects both viewing geometry and source-dependent emission physics. This work provides a framework for future Stokes-level and multi-epoch polarimetric studies with missions such as \textit{eXTP}.
\end{abstract}




\section{Introduction}
\label{sec:intro}
Magnetars are thought to be young isolated neutron stars with ultra-strong magnetic fields. These objects display a rich variety of X-ray phenomena, including bursts, flares, and timing irregularities such as glitches and anti- glitches~\citep{Kaspi2017ARAA}. The predominant energy source for their high-energy activity is thought to be the decay of their internal magnetic fields~\citep{Duncan1992ApJ}. To date, observations have confirmed or identified roughly 30 magnetars and candidates~\citep{Olausen2014ApJS}. Their persistent X-ray emission, bursting behavior, outbursts, and long-term variability are commonly interpreted as manifestations of coupled internal magnetic evolution and external magnetospheric restructuring \citep{Turolla2015RPPh, Kaspi2017ARAA}. In this context, the radiation geometry is directly connected with the orientation of the magnetic axis, the location of the emitting region, and the large-scale structure of the magnetosphere. However, integral observables such as spin period, spin-down rate, or phase-averaged luminosity alone generally do not determine the three-dimensional source geometry uniquely. A more direct observable is therefore required.

Polarization offers one of the most effective ways to access this information. The polarization signal is sensitive to the local magnetic-field orientation, to the emission mode, and to the subsequent propagation of photons through a strongly magnetized environment. In particular, the phase dependence of the polarization position angle (PA) retains information on how the projected field direction evolves as the star rotates, making it a natural tracer of the large-scale radiation geometry. For magnetars, this is especially valuable because their external fields may deviate substantially from a simple dipole through the presence of magnetospheric currents and twisted field lines \citep{Thompson2002ApJ,Beloborodov2009ApJ,Taverna2014MNRAS}.

At the same time, the interpretation of magnetar polarization must take into account strong-field propagation effects. Radiative-transfer calculations have shown that vacuum polarization can substantially modify the emerging polarization signal from highly magnetized neutron-star atmospheres, while magnetospheric propagation and scattering can further reshape the observed polarization degree and PA \citep{LaiHo2003ApJ,Taverna2014MNRAS,Taverna2022Sci}. X-ray polarization is therefore sensitive not only to the surface emission geometry, but also to the magnetospheric environment through which the radiation propagates.

A standard starting point for interpreting PA swings is the rotating vector model (RVM), originally developed for radio pulsars\,\citep{Radhakrishnan1969ApL}. In its classical form, the model assumes that the magnetic field is approximately dipolar and that the observed polarization direction follows the projected field orientation. For magnetars, however, the applicability of the classical RVM is less secure, because their magnetospheres are expected to be affected by large-scale currents, twists, and magnetic reconfiguration\,\citep{Thompson2002ApJ,Beloborodov2009ApJ}. Even in a dipolar geometry, finite emission height, aberration, and retardation can introduce additional shifts in the PA swing\,\citep{Blaskiewicz1991ApJ}.
Recent radio pulsar studies further show that PA curves and pulse profiles can depend on observing frequency, emission state, and single-pulse variability \citep{Wen2022ApJ,Wen2020ApJ,Wang2025ApJ,Wen2026ApJ,Yuen2024ApJ,Yuen2026MNRAS}, suggesting that RVM-derived geometries should generally be interpreted as effective, band-dependent constraints.
A magnetar-specific extension of the RVM has been proposed for twisted dipolar fields, providing a simple framework in which deviations of the PA swing from the classical dipole case can be related to magnetospheric twist\,\citep{Tong2021MNRAS}. In a twisted dipole, the azimuthal field component introduces an additional rotation of the projected field direction, thereby modifying the PA swing beyond the classical RVM prediction. It is therefore necessary to test whether the observed X-ray PA curves can already be explained within a dipole-dominated picture or whether a twisted-magnetosphere extension is statistically warranted. 

This problem has become observationally accessible with the advent of the \textit{Imaging X-ray Polarimetry Explorer} (\textit{IXPE}) \citep{Weisskopf2022JATIS}. Recent \textit{IXPE} observations have established X-ray polarization in magnetars and shown that phase-resolved polarimetry can provide direct constraints on their radiation geometry \citep{Taverna2022Sci,Heyl2024MNRAS}. These developments motivate a systematic comparison between geometrical models of different complexity using the available X-ray polarization data.

In this paper, we focus on two magnetars, 1E~2259+586 and 1E~1547.0$-$5408, both of which have published \textit{IXPE} phase-resolved polarimetric measurements suitable for geometrical modeling. These two sources show markedly different X-ray polarization properties: 1E~2259+586 is weakly polarized, whereas 1E~1547.0$-$5408 exhibits a much higher polarization degree and a well-defined phase-dependent PA swing. They therefore provide a useful pair for testing whether magnetar X-ray polarization diversity can be interpreted within a common geometrical framework.

The key question addressed in this work is whether the current phase-resolved X-ray PA data require a departure from a dipole-dominated geometry. Although magnetar magnetospheres are expected to contain currents and twisted field lines, it is not obvious whether such effects are statistically required by the available \textit{IXPE} measurements. We therefore compare two phenomenological geometrical descriptions: the classical rotating vector model (CRVM), which represents the dipole-dominated baseline, and a modified rotating vector model (MRVM), which includes a first-order correction associated with a globally twisted magnetosphere.

The main contribution of this work is a uniform Bayesian comparison between a dipole-dominated RVM and a twisted-magnetosphere extension for two IXPE magnetars with markedly different polarization properties. Building on the CRVM analysis of 1E~1547.0$-$5408 by \cite{Taverna2026arXiv}, we extend the comparison by including the MRVM, Bayesian evidence, and radio-informed priors based on the radio RVM geometry of \cite{Stewart2025arXiv}.
Rather than assuming that a twisted model is required, we quantify whether the additional twist parameter is statistically warranted by the current phase-resolved PA data.
Specifically, this work presents three main contributions: 
(i) a consistent Bayesian model-selection framework, including AIC, BIC, and Bayes factors, applied to the CRVM--MRVM comparison for two magnetars with contrasting polarization properties; 
(ii) a quantitative assessment of the radio--X-ray geometrical consistency for 1E~1547.0$-$5408 through radio-informed priors, showing that the nearly aligned radio geometry is compatible with, but not independently preferred by, the X-ray PA data; and 
(iii) a comparative characterization of the viewing geometries of the two sources, revealing different impact angles and supporting the view that magnetar X-ray polarization diversity reflects both viewing geometry and source-dependent emission physics.

The paper is organized as follows. Section~\ref{sec:data} describes the observational data. Section~\ref{sec:method} presents the geometrical models and Bayesian inference method. Section~\ref{sec:results} gives the fitting and model-comparison results. Section~\ref{sec:discussion} discusses the implications for magnetar radiation geometry, magnetospheric twist, and the origin of X-ray polarization diversity. Section~\ref{sec:summary} summarizes our main conclusions.

\section{Observational Data}
\label{sec:data}

We use published IXPE phase-resolved polarimetric measurements of two magnetars, 1E~2259+586 and 1E~1547.0$-$5408, to constrain their X-ray radiation geometries. IXPE measures the linear polarization degree (PD) and polarization position angle in the $2$--$8\,\mathrm{keV}$ band, providing a direct probe of viewing geometry and magnetic-field configuration in strongly magnetized neutron stars\,\citep{Weisskopf2022JATIS}. For magnetars, phase-resolved X-ray polarimetry is particularly useful because both surface emission and magnetospheric propagation effects, such as vacuum birefringence and resonant scattering, can imprint phase-dependent polarization signatures\,\citep{Taverna2014MNRAS,Taverna2024Galax}.

For 1E~2259+586, we adopt the phase-resolved IXPE results reported by\,\cite{Heyl2024MNRAS}. The source was observed from 2023 June 2 to June 19 and from 2023 June 30 to July 6, with a total exposure of approximately $1.2\,\mathrm{Ms}$. A phase-coherent timing solution was obtained by combining IXPE, NICER, and XMM--Newton data. In the $2$--$8\,\mathrm{keV}$ band, the phase-averaged polarization was detected with $\mathrm{PD}=(5.6\pm1.4)\%$, exceeding the corresponding $\mathrm{MDP}_{99}=4.5\%$, and with a mean $\mathrm{PA}=-75.2^\circ\pm7.4^\circ$. The rotational cycle was divided into 14 equal phase bins, for which the normalized Stokes parameters, PD, and PA were derived. The adopted data are listed in Table~\ref{tab:pa_phase_1e2259_heyl}. Since several phase bins have PD values below $\mathrm{MDP}_{99}$, their PAs are less robust; therefore, in the geometrical fits below, we use only the phase bins with PD values above $\mathrm{MDP}_{99}$. The retained bins are those with phase intervals $0.000$--$0.071$, $0.143$--$0.214$, $0.214$--$0.286$, $0.286$--$0.357$, $0.429$--$0.500$, $0.857$--$0.929$, and $0.929$--$1.000$.

\begin{table}[!htbp]
\caption{Phase-resolved polarization data of 1E~2259+586 in the $2$--$8\,\mathrm{keV}$ band.}
\label{tab:pa_phase_1e2259_heyl}
\centering
\footnotesize
\setlength{\tabcolsep}{5pt}
\renewcommand{\arraystretch}{1.15}
\begin{tabular}{ccccc}
\hline
Phase Interval & PA ($^\circ$) & $\sigma_{\rm PA}$ ($^\circ$) & PD (\%) & $\mathrm{MDP}_{99}$ (\%) \\
\hline
0.000--0.071 &  44.7  &  6.9  & $22.5\pm5.1$ & 16.3 \\
0.071--0.143 &  13.9  & 27.7  &  $4.5\pm4.1$ & 13.5 \\
0.143--0.214 & -75.8  &  8.1  & $13.6\pm3.6$ & 11.8 \\
0.214--0.286 & -71.6  &  6.3  & $17.5\pm3.7$ & 11.8 \\
0.286--0.357 & -66.1  &  4.9  & $22.5\pm3.9$ & 12.6 \\
0.357--0.429 & -65.4  &  9.8  & $13.1\pm4.4$ & 14.1 \\
0.429--0.500 & -65.6  &  6.1  & $22.2\pm5.0$ & 15.0 \\
0.500--0.571 & -47.9  & 10.5  & $12.8\pm4.9$ & 15.3 \\
0.571--0.643 &  14.2  & 57.0  &  $2.3\pm5.0$ & 14.5 \\
0.643--0.714 &  76.3  & 32.5  &  $2.9\pm4.5$ & 13.6 \\
0.714--0.786 &  -6.4  & 11.4  & $10.0\pm4.1$ & 12.6 \\
0.786--0.857 & -33.1  & 26.5  &  $4.5\pm4.2$ & 13.0 \\
0.857--0.929 &  39.7  &  7.2  & $19.7\pm5.0$ & 15.6 \\
0.929--1.000 &  53.1  &  5.8  & $25.7\pm5.3$ & 16.5 \\
\hline
\end{tabular}
\end{table}

For 1E~1547.0$-$5408, we use the recent IXPE phase-resolved polarization measurements presented by\,\cite{Taverna2026arXiv}. The observation was carried out from 2025 March 26 to April 5, with a total effective exposure of approximately $500\,\mathrm{ks}$. In the $2$--$6\,\mathrm{keV}$ band, this source shows a much stronger phase-averaged polarization signal than 1E~2259+586, with $\mathrm{PD}=(47.7\pm2.9)\%$ and mean $\mathrm{PA}=75.8^\circ\pm1.8^\circ$. The original analysis divided one rotational cycle into seven equal phase intervals and found a clear, continuous variation of the PA with phase. Since linear polarization angles are defined modulo $180^\circ$, we adopt the unfolded PA branch used for geometrical modeling, in which the third and fourth phase bins are shifted by subtracting $180^\circ$. The original measured values (before unfolding) were approximately $86^\circ$ and $88^\circ$, yielding the unfolded values of $-94^\circ$ and $-92^\circ$ listed in Table~\ref{tab:pa_phase_1e1547}. All subsequent modeling uses these unfolded values. 

\begin{table}[!htbp]
\caption{Phase-resolved polarization position angle data of 1E~1547.0$-$5408 in the $2$--$6\,\mathrm{keV}$ band.}
\label{tab:pa_phase_1e1547}
\centering
\footnotesize
\setlength{\tabcolsep}{7pt}
\renewcommand{\arraystretch}{1.2}
\begin{tabular}{cccc}
\hline
Phase Interval & Phase Center & PA ($^\circ$) & $\sigma_{\rm PA}$ ($^\circ$) \\
\hline
0.00--0.14 & 0.070 & -63.6 & 5.2 \\
0.14--0.29 & 0.215 & -79.2 & 3.6 \\
0.29--0.43 & 0.360 & -94.0 & 5.4 \\
0.43--0.57 & 0.500 & -92.0 & 4.7 \\
0.57--0.71 & 0.640 & -82.3 & 3.4 \\
0.71--0.86 & 0.785 & -65.6 & 3.6 \\
0.86--1.00 & 0.930 & -48.1 & 3.9 \\
\hline
\end{tabular}
\end{table}

In addition to the X-ray polarization data, 1E~1547.0$-$5408 has an independent radio geometrical constraint. It is one of the few magnetars with transient pulsed radio emission. Based on Parkes/Murriyang observations in the $2.5$--$4.0\,\mathrm{GHz}$ band, \citet{Stewart2025arXiv} fitted the radio polarization data with the rotating vector model and obtained $\chi_{\rm r}=3.4^{+1.3}_{-1.2}\,\mathrm{deg}$ and $\zeta_{\rm r}=7.5^{+3.0}_{-2.6}\,\mathrm{deg}$, suggesting a nearly aligned geometry. We use these radio constraints as an external reference for the X-ray-only solutions and, in a separate set of fits, as informative priors to test whether the X-ray PA curve is compatible with the same large-scale geometry inferred from radio polarization. This allows us to evaluate both the independent constraining power of the X-ray data and the sensitivity of the inferred geometry to prior information.

The PA uncertainties in both datasets are assumed to be Gaussian in our likelihood analysis (Section~\ref{sec:method}). This approximation is justified for bins with significant polarization detection, i.e., where the PD is well above the $\mathrm{MDP}_{99}$. For 1E~2259+586, only bins with $\mathrm{PD} > \mathrm{MDP}_{99}$ are retained; for 1E~1547.0$-$5408, all bins have high PD and consequently reliable Gaussian uncertainties. Sensitivity tests confirm that excluding the borderline bins (e.g., the $0.143$--$0.214$ bin of 1E~2259+586, whose PD is only marginally above $\mathrm{MDP}_{99}$) does not alter our main conclusions.

\section{Geometrical Modeling and Bayesian Analysis}
\label{sec:method}

\subsection{The Rotating Vector Model and its Extension}
In the ultra-strong magnetic environment surrounding a magnetar, radiation propagates in two normal polarization modes, namely the ordinary (O) mode and the extraordinary (X) mode. In the former, the electric vector oscillates in the plane defined by the local magnetic field and the photon wave vector, whereas in the latter it oscillates perpendicular to that plane\,\citep{Gnedin1978SvAL, Pavlov1979JETP}. Owing to vacuum birefringence, the polarization vector is expected to follow the direction of the local magnetic field adiabatically until the so-called polarization-limiting radius\,\citep{Heyl2000MNRAS, Heyl2002PRD}. For keV photons from typical magnetars, this radius is expected to be of the order of a few hundred stellar radii, where the large-scale field is dominated by its dipolar component\,\citep{Taverna2015MNRAS, HeylCaiazzo2018Galaxies}. As a result, the polarization measured by a distant observer is expected to be either parallel or perpendicular to the instantaneous projection of the magnetic axis onto the plane of the sky, depending on whether the escaping radiation is dominated by the O or X mode. Under these conditions, the phase modulation of the polarization angle is largely decoupled from the evolution of the polarization degree and intensity, and is expected to trace mainly the viewing geometry. This makes the rotating vector model a natural starting point for describing the phase dependence of the polarization angle in magnetars\,\citep{Radhakrishnan1969ApL, Poutanen2020AA, Taverna2022Sci, GonzalezCaniulef2023MNRAS}.

We model the phase-resolved polarization position angle curves of both sources using two geometrical descriptions: the CRVM and MRVM. The CRVM provides the baseline dipole-dominated geometry, while the MRVM introduces a first-order correction associated with a globally twisted magnetosphere.

For the CRVM, we adopt the standard expression\,\citep{Radhakrishnan1969ApL, LaiHo2002ApJ}. 
\begin{equation}
\Psi_{\rm CRVM}(\phi)=\Psi_{0}+\arctan\!\left[
\frac{-\sin\chi\,\sin(\phi-\phi_{0})}
{\sin\zeta\cos\chi-\cos\zeta\sin\chi\cos(\phi-\phi_{0})}
\right],
\label{eq:crvm}
\end{equation}
where $\Psi$ represents the position angle, $\Psi_0$ is the position angle of the pulsar spin axis, $\chi$ is the magnetic inclination angle, $\zeta$ is the angle between the pulsar spin vector and the line of sight, $\phi$ is the pulse phase, and $\phi_0/2\pi$ is the phase (in units of the rotational cycle) at which the magnetic pole is closest to the observer.

To account for possible magnetospheric twist, we further consider the MRVM proposed by \citet{Tong2021MNRAS}. In this approximation, the observed PA is written as
\begin{equation}
\Psi_{\rm MRVM}(\phi)=\Psi_{\rm CRVM}(\phi)+\Delta\Psi_{\rm twist}(\phi),
\label{eq:mrvm_basic}
\end{equation}
where the leading-order twisted contribution is approximated by
\begin{equation}
\Delta\Psi_{\rm twist}\simeq -\frac{8}{9}\lambda \sin^{2}\theta_{\rm obs},
\label{eq:dpsi_twist}
\end{equation}
where $\theta_{\rm obs}$ is the colatitude of the line of sight in the magnetic frame which is $\cos\theta_{\rm obs}=
\cos\chi\cos\zeta+\sin\chi\sin\zeta\cos(\phi-\phi_{0})$.
Here, $\lambda$ is a dimensionless parameter that parametrizes the twist of the magnetic field through $\lambda=\sqrt{(35/16)(1-n)}$, where $n=1$ corresponds to the pure dipole limit, $n=0$ is the split monopole case, and $0<n<1$ is the general twisted dipole case.

Equation~(\ref{eq:dpsi_twist}) is the leading-order correction valid for small twist amplitudes; for $\lambda$ approaching the upper bound of $1.48$, higher-order terms may become non-negligible, but the present data do not require them (see Section~\ref{sec:discussion}). In the present work, we use this first-order correction as a phenomenological extension of the CRVM. Its role is not to provide a complete description of the magnetar magnetosphere, but to test whether the available X-ray PA data require a statistically significant departure from the dipole-dominated case.

For both models, we also report the derived impact angle $\beta=\zeta-\chi$, which measures the minimum angular separation between the line of sight and the magnetic axis.

\subsection{Bayesian inference method}
\label{sec:bayesian}

The phase-resolved PA data of 1E~2259+586 and 1E~1547.0$-$5408 are analyzed within a common Bayesian framework. For both sources, we first perform a baseline analysis using uninformative priors. For 1E~1547.0$-$5408, we additionally carry out a radio-informed analysis by incorporating the radio-derived geometrical constraints as informative priors. This allows us to examine whether the X-ray PA curve is compatible with the nearly aligned geometry inferred from radio polarization and to assess the prior sensitivity of the X-ray geometrical inference.
The parameter vectors of the two models are $\boldsymbol{\Theta}_{\rm CRVM}=(\Psi_{0},\chi,\zeta,\phi_{0}/2\pi)$,
and $\boldsymbol{\Theta}_{\rm MRVM}=(\Psi_{0},\chi,\zeta,\phi_{0}/2\pi,\lambda)$.

Because the PA is an axial quantity defined modulo $180^{\circ}$, a naive linear residual can introduce artificial discontinuities near branch boundaries. We therefore compute the residual using the standard double-angle representation of linear polarization
\citep{Naghizadeh1993AA,Everett2001ApJ}:
\begin{equation}
\Delta_{i}
=
\frac{1}{2}\operatorname{atan2}\!\left[
\sin(2\delta_{i}),
\cos(2\delta_{i})
\right],
\qquad
\delta_{i}={\rm PA}_{{\rm obs},i}-{\rm PA}_{{\rm mod},i},
\label{eq:angular_residual}
\end{equation}
where ${\rm PA}_{{\rm obs},i}$ and ${\rm PA}_{{\rm mod},i}$ are the observed PA and the model prediction.
This definition ensures that $\Delta_i$ is always wrapped into the physically relevant range. Although some PA curves are displayed in an unwrapped form for clarity, all statistical inferences are performed using this periodic angular residual.

In principle, the most fundamental treatment of X-ray polarimetric data is to fit the Stokes parameters directly. In the present work, however, we use the published phase-resolved PA measurements and their reported uncertainties, rather than reprocessing the \textit{IXPE} event files. We therefore adopt a PA-level likelihood as a practical geometrical description of the data. This approximation is most appropriate when the polarization detection in a phase bin is significant and the PA uncertainty is approximately Gaussian. For 1E~2259+586, we mitigate the possible non-Gaussian behavior of poorly constrained PA measurements by using only the phase bins with PD above $\mathrm{MDP}_{99}$. Nevertheless, for bins only marginally above this threshold, the PA likelihood may still deviate from a purely Gaussian form because the underlying polarization amplitude is measured with finite signal-to-noise. A likelihood formulated directly in terms of the Stokes parameters would avoid this PA-level approximation. For 1E~1547.0$-$5408, the much higher polarization degree makes the phase-resolved PA measurements more robust.

Following the standard Gaussian-error likelihood commonly used in astronomical parameter estimation\,\citep{Lampton1976ApJ,Hogg2010arXiv},
and assuming independent measurement uncertainties, the log-likelihood is
written as
\begin{equation}
\ln \mathcal{L}
=
-\frac{1}{2}\sum_{i}
\left[
\frac{\Delta_{i}^{2}}{\sigma_{{\rm PA},i}^{2}}
+
\ln\!\left(2\pi\sigma_{{\rm PA},i}^{2}\right)
\right],
\label{eq:loglike}
\end{equation}
where $\sigma_{{\rm PA},i}$ is the uncertainty of the observed PA in the $i$th phase bin.

For the baseline analysis, we adopt uniform priors
$\Psi_{0}\in(-180^{\circ},\,180^{\circ}),\,
\chi\in(0^{\circ},\,90^{\circ}),\,
\zeta\in(0^{\circ},\,180^{\circ}),\,
\phi_{0}/2\pi\in(-0.5,\,0.5)$.
For the MRVM, we further impose
$\lambda\in(0,\,1.48)$,
which corresponds to the physically relevant range $0<n<1$. The posterior distribution is therefore
\begin{equation}
p(\boldsymbol{\Theta}\mid\mathcal{D})
\propto
\mathcal{L}(\mathcal{D}\mid\boldsymbol{\Theta})\,\pi(\boldsymbol{\Theta}),
\label{eq:posterior}
\end{equation}
where $\mathcal{D}$ denotes the phase-resolved PA data and $\pi(\boldsymbol{\Theta})$ is the prior. For 1E~1547.0$-$5408, we additionally perform a second analysis using
Gaussian priors for $\chi$ and $\zeta$ informed by the radio constraints,
$\chi \sim \mathcal{N}(3.4^{\circ},1.3^{\circ})$ and
$\zeta \sim \mathcal{N}(7.5^{\circ},3.0^{\circ})$
\citep{Stewart2025arXiv}. The remaining parameters are assigned the same
priors as in the baseline analysis.

\subsection{Sampling Strategy and Model Comparison}

We first use differential evolution to identify a high-likelihood region in parameter space \citep{Storn1997JGOpt}. The best solution obtained in this stage is then used to initialize an ensemble Markov chain Monte Carlo sampler based on \texttt{emcee} \citep{ForemanMackey2013PASP}. Posterior summaries are reported using the median together with the 16th and 84th percentiles, which we quote as the central value and the corresponding $1\sigma$ credible interval.

To estimate the Bayesian evidence, we also perform nested-sampling calculations using \texttt{dynesty}\,\citep{Speagle2020MNRAS}. This allows us to compare the global statistical support for the CRVM and MRVM rather than relying only on best-fitting solutions.

We compare the CRVM and MRVM using both information criteria and Bayesian evidence. Specifically, we compute the Akaike information criterion (AIC) and Bayesian information criterion (BIC), which penalize model complexity in addition to goodness of fit\,\citep{Akaike1974ITAC,Schwarz1978AnSta}. We also evaluate the Bayesian evidence difference between the two models and quote the corresponding Bayes factor, with positive values favoring the MRVM and negative values favoring the CRVM\,\citep{Kass1995JASA,Trotta2008ConPh}. Finally, posterior predictive distributions are generated from the posterior samples to visualize the range of PA curves allowed by each model and to identify phase intervals where the CRVM and MRVM make distinguishable predictions\,\citep{Gabry2017arXiv}.

\section{Results}
\label{sec:results}

\subsection{Results for 1E~2259+586}
\label{sec:2259_results}

We first fit the phase-resolved PA data of 1E~2259+586 in the $2$--$8\,\mathrm{keV}$ band using the CRVM and MRVM. The posterior constraints on the geometrical parameters are summarized in Table~\ref{tab:2259_post}. The corresponding posterior distributions are shown in Figure~\ref{fig:app_corner_2259}.

\begin{table}[h]
\caption{Posterior constraints on the geometrical parameters of 1E~2259+586.}
\label{tab:2259_post}
\centering
\footnotesize
\setlength{\tabcolsep}{10pt}
\renewcommand{\arraystretch}{1.5}
\begin{tabular}{lccccc}
\hline
Model & $\Psi_0$ ($^\circ$) & $\chi$ ($^\circ$) & $\zeta$ ($^\circ$) & $\phi_0/2\pi$ & $\lambda$ \\
\hline
CRVM
& $-13.1^{+2.7}_{-2.6}$
& $53.7^{+5.6}_{-7.2}$
& $64.6^{+8.8}_{-10.4}$
& $0.120^{+0.020}_{-0.020}$
& -- \\
MRVM
& $-4.30^{+5.7}_{-5.1}$
& $30.8^{+10.9}_{-6.9}$
& $35.7^{+13.7}_{-8.6}$
& $0.070^{+0.020}_{-0.020}$
& $0.86^{+0.40}_{-0.43}$ \\
\hline
\end{tabular}
\end{table}
For the CRVM, the inferred geometry is characterized by a moderate magnetic inclination and a moderate viewing angle. The corresponding impact angle is
$\beta\simeq 11^\circ$.
For the MRVM, the posterior median gives
$\beta\simeq 5^\circ$.
Although the two models yield different central values for $\chi$ and $\zeta$, both favor a relatively small impact angle, implying that the line of sight passes close to the magnetic axis during the rotation cycle.

The twist parameter in the MRVM is constrained to $\lambda=0.86^{+0.40}_{-0.43}$. However, the posterior distribution is broad, indicating that the current data do not tightly determine the twist strength. This also reflects a degeneracy between the geometrical angles and the twist correction: the additional degree of freedom in the MRVM improves the flexibility of the fit, but weakens the independent constraints on $\chi$ and $\zeta$.

To further assess the relative performance of the two geometrical models, we calculate the goodness of fit at the maximum-likelihood point, the information criteria, and the Bayesian evidence. The results are summarized in Table~\ref{tab:model_compari_2259}.

\begin{table}[h]
\centering
\caption{Model comparison results for 1E~2259+586.}
\label{tab:model_compari_2259}
\footnotesize
\setlength{\tabcolsep}{10pt}
\renewcommand{\arraystretch}{1.5}
\begin{tabular}{lccccc}
\hline
Model & $\chi^2_{\min}$ & $\ln \mathcal{L}_{\max}$ & AIC & BIC & $\ln Z$ \\
\hline
CRVM & 7.89 & -23.37 & 54.74 & 54.53 & -34.52 \\
MRVM & 1.81 & -20.33 & 50.66 & 50.39 & -33.34 \\
\hline
\end{tabular}
\end{table}

\begin{figure*}[h]
\centering
\includegraphics[width=0.7\textwidth]{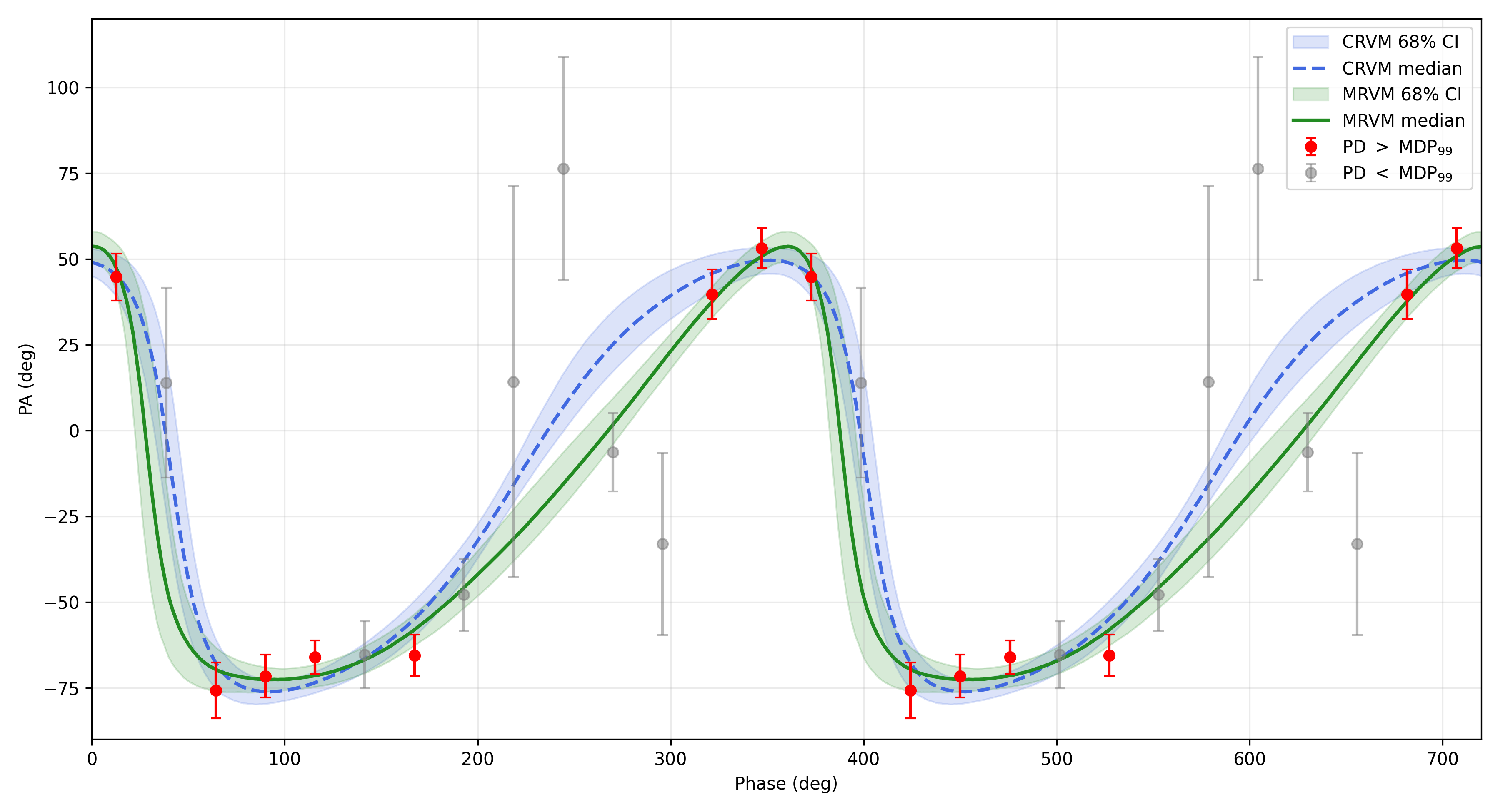}
\caption{
Posterior predictive distributions for the phase-resolved PA curve of 1E~2259+586.
The red points show the IXPE measurements used in the fit.
The blue and green curves denote the posterior median predictions of the CRVM and MRVM, respectively, while the corresponding shaded regions show the 68\% credible intervals.
For clarity, the model curves are displayed over two rotational cycles.
}
\label{fig:ppd_2259}
\end{figure*}

As shown in Table~\ref{tab:model_compari_2259}, the MRVM achieves a smaller minimum chi-square than the CRVM, indicating that the additional twist parameter improves the description of the observed PA curve at the maximum-likelihood level. The MRVM also gives lower AIC and BIC values, suggesting that the improvement in fit quality is not completely offset by the penalty for the extra model parameter. The evidence difference is
$\Delta \ln Z=1.18$,
which corresponds to a Bayes factor
${\rm BF}=\exp(\Delta\ln Z)=3.25$.
This indicates a mild preference for the MRVM over the CRVM. However, the evidence is still not strong enough to claim that the current IXPE data require a statistically significant global twist of the magnetosphere.

The posterior predictive distributions in Figure~\ref{fig:ppd_2259} show the same trend. Both models reproduce the overall phase evolution of the PA, while the MRVM provides a somewhat more flexible description near the rapidly varying part of the curve. Nevertheless, the two predictive bands overlap over most of the phase range, and the improvement introduced by the twist correction remains moderate.

Taken together, the posterior constraints and model comparison results suggest that the phase modulation of the PA in 1E~2259+586 is primarily governed by the large-scale viewing geometry. The CRVM already provides an acceptable description of the data, while the inclusion of a twist-induced correction improves the fit only moderately. Therefore, based on the single-epoch phase-resolved IXPE data alone, a twisted magnetosphere remains possible but is not statistically required.

Both models also imply a relatively small impact angle. Although the inferred values of $\chi$ and $\zeta$ differ between the CRVM and MRVM, the line of sight passes close to the magnetic axis in both solutions. This small-impact-angle configuration is useful for interpreting the smooth phase-dependent PA swing of 1E~2259+586, and it provides an important contrast with the geometry inferred for 1E~1547.0$-$5408 below.

\subsection{Results for 1E~1547.0$-$5408}
\label{sec:1547_results}

\subsubsection{Constraints from the X-ray Data with Flat Priors}
\label{subsubsec:1547_flat}

\begin{figure*}[!htbp]
\centering
\includegraphics[width=0.7\textwidth]{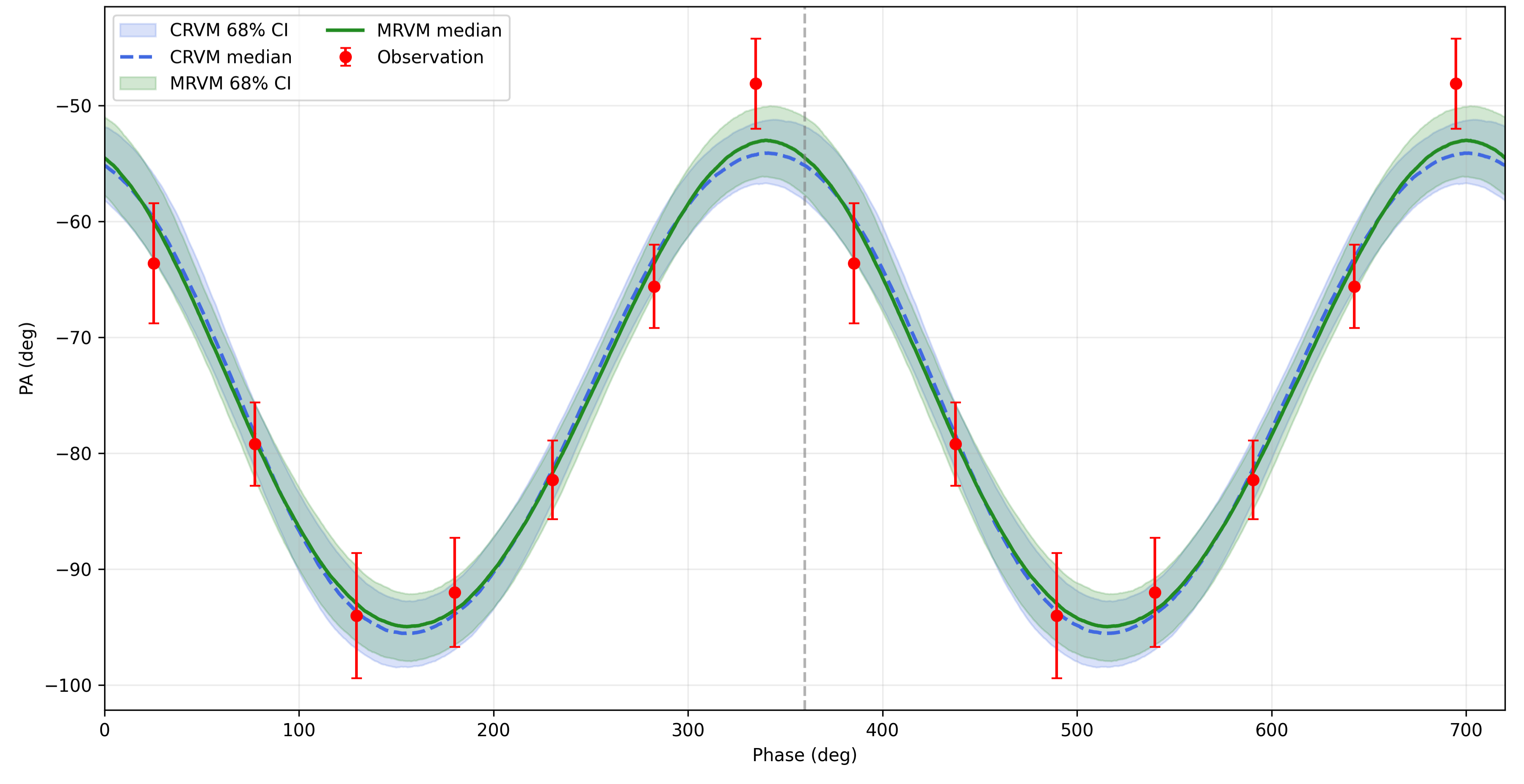}
\caption{
Posterior predictive distributions for the phase-resolved PA curve of 1E~1547.0$-$5408 obtained with flat priors.
The red points show the IXPE measurements used in the fit.
The blue dashed and green solid curves denote the posterior median predictions of the CRVM and MRVM, respectively.
The blue and green shaded regions show the corresponding 68\% credible intervals.
For clarity, the model curves are displayed over two rotational cycles.
}
\label{fig:ppd_1547_flat}
\end{figure*}

We fit the $2$--$6\,\mathrm{keV}$ phase-resolved PA data of 1E~1547.0$-$5408 using the CRVM and MRVM under flat priors. This fit is treated as the X-ray-only baseline, against which the radio-informed analysis below will be compared. The posterior constraints are summarized in Table~\ref{tab:1547_post}, and the corresponding posterior distributions are shown in Figure~\ref{fig:app_corner_1547_flat}.
The corresponding model comparison results are listed in Table~\ref{tab:1547_cmp}, and the posterior predictive distribution is shown in Figure~\ref{fig:ppd_1547_flat}.

\begin{table}[h]
\centering
\caption{Posterior constraints on the geometrical parameters of 1E~1547.0$-$5408 with flat priors.}
\label{tab:1547_post}
\footnotesize
\setlength{\tabcolsep}{10pt}
\renewcommand{\arraystretch}{1.5}
\begin{tabular}{lccccc}
\hline
Model & $\Psi_0$ ($^\circ$) & $\chi$ ($^\circ$) & $\zeta$ ($^\circ$) & $\phi_0/2\pi$ & $\lambda$  \\
\hline
CRVM
& $-74.8^{+1.5}_{-1.6}$
& $19.1^{+3.0}_{-4.4}$
& $84^{+29}_{-29}$
& $0.19^{+0.02}_{-0.02}$
& -- \\
MRVM
& $127^{+19}_{-15}$
& $14.0^{+4.7}_{-3.7}$
& $71^{+62}_{-30}$
& $0.18^{+0.10}_{-0.08}$
& $0.87^{+0.43}_{-0.59}$\\
\hline
\end{tabular}
\end{table}

As shown in Table~\ref{tab:1547_post}, the CRVM already provides stable constraints on the phase-resolved PA curve of 1E~1547.0$-$5408. The posterior median corresponds to a relatively small magnetic inclination and a large viewing angle. The corresponding impact angle is
$\beta\simeq 64^\circ$ ,
which is much larger than that inferred for 1E~2259+586. This suggests that, for 1E~1547.0$-$5408, the line of sight does not pass close to the magnetic axis, but instead samples the projected radiation geometry at a relatively large angular separation from the magnetic pole.

The constraint on $\zeta$ is significantly broader than that on $\chi$, indicating a remaining degeneracy in the absolute viewing angle. In particular, the posterior distribution in the $\chi$--$\zeta$ plane follows an extended correlated region, implying that the observed PA curve mainly constrains the swing amplitude and phase reference, while the decomposition into individual geometrical angles is less unique. Therefore, the most robust result from the X-ray-only fit is that the magnetic inclination is small, whereas the viewing angle is only weakly constrained toward relatively large values.

Compared with the CRVM, the MRVM gives a similar qualitative geometry, still favoring a small $\chi$ and a relatively large $\zeta$. However, the uncertainties become broader, especially for $\zeta$ and $\lambda$. This indicates a strong coupling between the twist correction and the geometrical parameters. Although the posterior allows nonzero values of $\lambda$, the broad distribution of the twist parameter shows that the current X-ray PA data do not provide a tight constraint on the magnetospheric twist. In this sense, the MRVM should be regarded as a flexible extension of the CRVM rather than as evidence for a required global twist.

\begin{table}[h]
\centering
\caption{Model comparison results for 1E~1547.0$-$5408 with flat priors.}
\label{tab:1547_cmp}
\footnotesize
\setlength{\tabcolsep}{10pt}
\renewcommand{\arraystretch}{1.5}
\begin{tabular}{lccccc}
\hline
Model & $\chi^2_{\min}$ & $\ln \mathcal{L}_{\max}$ & AIC & BIC & $\ln Z$ \\
\hline
CRVM
& 3.582
& -18.253
& 44.51
& 44.29
& -29.18 \\
MRVM
& 3.584
& -18.254
& 46.51
& 46.24
& -28.92 \\
\hline
\end{tabular}
\end{table}

The quantitative comparison in Table~\ref{tab:1547_cmp} shows that the CRVM and MRVM describe the data almost equally well at the maximum-likelihood level. The minimum chi-square values are nearly identical. After accounting for the additional twist parameter, both AIC and BIC favor the CRVM. The Bayes factor is ${\rm BF}=1.29$.
This represents only a negligible preference for the MRVM in terms of evidence and is far from sufficient to claim that the twisted-magnetosphere model is statistically favored.

The same conclusion is supported by the posterior predictive distributions. The CRVM and MRVM median curves are nearly indistinguishable over the full phase range, and their 68\% credible intervals largely overlap. The observed PA curve is therefore well described as a smooth periodic swing, without requiring additional local distortions or sharp phase-dependent features. We thus regard the CRVM as an adequate description of the X-ray PA data of 1E~1547.0$-$5408 under flat priors.

\subsubsection{Radio-informed Cross-check}
\label{subsubsec:1547_radio}

\begin{figure*}[!htbp]
\centering
\includegraphics[width=0.82\textwidth]{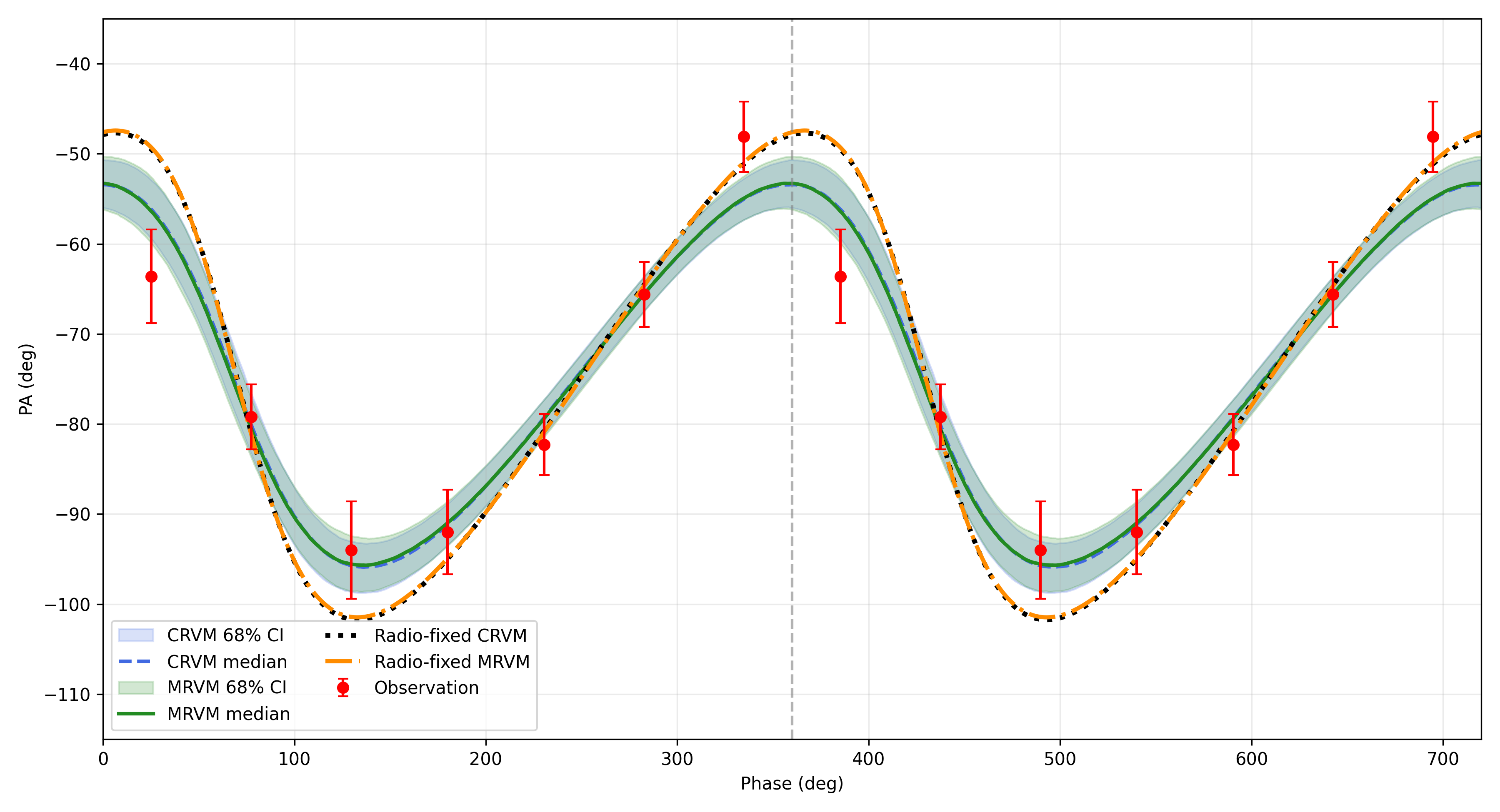}
\caption{
Posterior predictive distributions for the phase-resolved PA curve of
1E~1547.0$-$5408 obtained with radio-informed priors. The red points
show the IXPE PA measurements in the 2--6 keV band. The blue dashed
and green solid curves denote the posterior median predictions of the
CRVM and MRVM, respectively, and the shaded regions show the
corresponding 68\% credible intervals. The black dotted and orange
dash-dotted curves show the radio-fixed CRVM and MRVM fits, respectively,
in which the geometrical angles are fixed to the radio RVM central values
of \citet{Stewart2025arXiv}, $\chi=3.4^\circ$ and $\zeta=7.5^\circ$. For clarity, the
model curves are displayed over two rotational cycles.
}
\label{fig:ppd_1547_radio}
\end{figure*}

Radio polarimetric observations of 1E~1547.0$-$5408 suggest a nearly aligned geometry, with $\chi_{\rm r}=3.4^{+1.3}_{-1.2}\,\mathrm{deg}$ and $\zeta_{\rm r}=7.5^{+3.0}_{-2.6}\,\mathrm{deg}$\,\citep{Stewart2025arXiv}. To examine whether the X-ray PA data are compatible with this radio geometry, and to quantify the sensitivity of the X-ray geometrical inference to external information, we perform an additional set of fits using Gaussian priors on $\chi$ and $\zeta$ based on the radio RVM constraints. This radio-informed analysis is used as a cross-check and should not be regarded as replacing the X-ray-only baseline obtained with flat priors. The resulting posterior constraints are summarized in Table~\ref{tab:post_1e1547_radio}, and the corresponding posterior distributions are shown in Figure~\ref{fig:app_corner_1547_radio}.
The corresponding posterior predictive distributions are shown in Figure~\ref{fig:ppd_1547_radio}.

\begin{table}[h]
\centering
\caption{Posterior constraints on the geometrical parameters of 1E~1547.0$-$5408 with radio-informed priors.}
\label{tab:post_1e1547_radio}
\footnotesize
\setlength{\tabcolsep}{6pt}
\renewcommand{\arraystretch}{1.15}
\begin{tabular}{lccccc}
\hline
Model & $\Psi_0$ ($^\circ$) & $\chi$ ($^\circ$) & $\zeta$ ($^\circ$) & $\phi_0/2\pi$ & $\lambda$ \\
\hline
CRVM
& $-74.7^{+1.5}_{-1.5}$
& $3.17^{+0.80}_{-0.77}$
& $8.8^{+2.2}_{-2.1}$
& $0.190^{+0.010}_{-0.020}$
& -- \\
MRVM
& $-73.5^{+1.8}_{-1.7}$
& $3.20^{+0.80}_{-0.76}$
& $8.9^{+2.2}_{-2.1}$
& $0.180^{+0.020}_{-0.020}$
& $0.78^{+0.48}_{-0.52}$ \\
\hline
\end{tabular}
\end{table}

As expected, the radio-informed priors pull the posterior distributions of $\chi$ and $\zeta$ toward the nearly aligned configuration. For the CRVM, we obtain $\chi=3.17^{+0.80}_{-0.77}\,\mathrm{deg}$ and $\zeta=8.78^{+2.16}_{-2.09}\,\mathrm{deg}$. For the MRVM, the corresponding constraints are $\chi=3.20^{+0.80}_{-0.76}\,\mathrm{deg}$ and $\zeta=8.92^{+2.18}_{-2.11}\,\mathrm{deg}$. Thus, once the radio prior is imposed, the two X-ray models converge to almost the same nearly aligned geometry.

However, the fit quality degrades compared with the flat-prior case. The minimum chi-square increases from 3.58 to 6.19 for the CRVM, and from 3.58 to 5.86 for the MRVM. This indicates that the nearly aligned radio geometry is not ruled out by the X-ray PA curve, but it is not the geometry independently preferred by the X-ray data alone. In this sense, the X-ray and radio constraints are consistent but not identical: the X-ray-only fit favors a small-$\chi$ and larger-$\zeta$ solution, while the radio-informed fit is driven toward the externally imposed radio geometry.

The radio-informed MRVM fit does not provide a statistically significant improvement over the CRVM. The twist parameter remains broadly constrained, $\lambda=0.78^{+0.48}_{-0.52}$, and the predicted PA curves are very similar to those of the CRVM. Therefore, even under the nearly aligned radio geometry, the X-ray data do not require a significant global magnetospheric twist.

To further visualize the comparison with the radio geometry, we also perform radio-fixed fits in which the geometrical angles are fixed to the radio RVM central values of \citet{Stewart2025arXiv}, $\chi=3.4^\circ$ and $\zeta=7.5^\circ$. For the radio-fixed CRVM, only $\Psi_0$ and $\phi_0$ are refitted, while for the radio-fixed MRVM, $\Psi_0$, $\phi_0$, and $\lambda$ are refitted. The resulting curves are overplotted in Figure~\ref{fig:ppd_1547_radio}. The radio-fixed CRVM gives $\chi^2_{\min}=10.91$, while the radio-fixed MRVM gives $\chi^2_{\min}=10.90$ with $\lambda=0.64$. The two radio-fixed curves are nearly indistinguishable, and the improvement in $\chi^2$ is negligible. 
Moreover, the information criteria disfavor the additional twist parameter in the radio-fixed case, with $\Delta{\rm AIC}=1.99$ and $\Delta{\rm BIC}=1.93$ for MRVM minus CRVM. Thus, fixing the geometry to the radio RVM central values gives a poorer description of the X-ray PA curve than the radio-informed fits, and the first-order twist correction does not provide a meaningful improvement under this fixed geometry.

Overall, the X-ray data independently favor a small magnetic inclination for 1E~1547.0$-$5408, which is qualitatively consistent with the radio result. However, the X-ray data alone do not require the extremely aligned configuration preferred by the radio fit. The systematic difference between the X-ray-only and radio-informed solutions suggests that the inferred geometry may be affected by waveband-dependent emission regions, propagation effects, or residual model degeneracies. A fully self-consistent joint fit to the radio and X-ray polarization data will be needed to determine whether both wavebands trace the same large-scale magnetic geometry.

\subsection{Comparison with Previous RVM Analyses of 1E~1547.0$-$5408}
\label{subsec:comparison_previous_1547}

It is useful to compare our 1E~1547.0$-$5408 results with the recent analyses of \citet{Taverna2026arXiv} and \citet{Stewart2025arXiv}, which used the same IXPE observing campaign but adopted different analysis strategies and physical emphases. \citet{Taverna2026arXiv} fitted the phase-dependent IXPE PA curve with the classical RVM, using the PA measurements in the 2--3 and 3--6 keV bands simultaneously. Their two equivalent RVM branches correspond to an inclined X-ray geometry with $(\chi,\zeta)\simeq(22^\circ,73^\circ)$, or equivalently $(158^\circ,107^\circ)$ under the opposite RVM branch convention.

Our flat-prior CRVM result is broadly consistent with this X-ray RVM geometry. We obtain $\chi\simeq19^\circ$ for the magnetic inclination and $\zeta\simeq84^\circ$ for the viewing angle. Thus, both analyses favor a small-to-moderate magnetic inclination and a large line-of-sight angle, rather than the nearly aligned radio geometry. The remaining quantitative difference in $\zeta$ is not unexpected, because our fit uses the 2--6 keV PA values in seven phase bins and a PA-level Bayesian likelihood, whereas \citet{Taverna2026arXiv} fitted the PA measurements in two energy bands and discussed the two equivalent RVM sign conventions. Our main extension beyond their work is that we also test the MRVM and perform an explicit Bayesian model comparison. The result that the MRVM does not improve the fit shows that the RVM-like X-ray PA swing identified by \citet{Taverna2026arXiv} does not require an additional global twist correction.

\citet{Stewart2025arXiv} reached a somewhat different conclusion. Their radio RVM fit to simultaneous Parkes/Murriyang data gives a nearly aligned geometry, with $\chi=3.4^{+1.3}_{-1.2}\,\deg$ and $\zeta=7.5^{+3.0}_{-2.6}\,\deg$, while their X-ray PA modeling gives $\chi=18^\circ\pm4^\circ$ and $\zeta=69^{+25}_{-23}\,\deg$. Our flat-prior CRVM solution agrees well with their X-ray magnetic inclination, although our viewing angle is larger. This difference can plausibly be attributed to the different PA data products, phase binning, likelihood choices, and priors used in the two analyses. More importantly, our radio-informed fits, in which the nearly aligned radio geometry of \citet{Stewart2025arXiv} is imposed as an informative prior, show that this geometry is not excluded by the X-ray PA data, but is not independently preferred by them. In our fits, the minimum chi-square increases from 3.58 to 6.19 for the CRVM and from 3.58 to 5.86 for the MRVM when the radio-informed priors are adopted.

Therefore, our results reproduce the qualitative radio--X-ray geometrical difference discussed in previous studies. They also show that introducing the first-order MRVM twist correction does not remove this difference. This conclusion does not contradict the vacuum-birefringence interpretation discussed by \citet{Stewart2025arXiv}, because their strongest evidence relies on Stokes-parameter and radiative-transfer modeling, whereas our analysis is restricted to the geometrical information encoded in the PA curve. In this sense, the present CRVM/MRVM comparison is complementary to previous Stokes-level and radiative-transfer studies.

\section{Discussion}
\label{sec:discussion}

\subsection{Do the IXPE Data Require a Twisted Magnetosphere?}
\label{subsec:discussion_twist}

The MRVM was introduced to test whether the phase-resolved X-ray PA curves require a measurable departure from the classical dipole-dominated geometry. For both magnetars studied here, we find no statistically compelling evidence for such a departure. The CRVM already provides an acceptable description of the observed PA modulation, whereas the MRVM gives only a modest improvement for 1E~2259+586 and essentially no improvement for 1E~1547.0$-$5408.

For 1E~2259+586, the MRVM gives a lower $\chi^2_{\min}$ and smaller AIC and BIC values than the CRVM. The Bayesian evidence also mildly favors the MRVM, with a Bayes factor of ${\rm BF}=3.25$. According to the scale of\,\cite{Kass1995JASA}, this constitutes positive evidence for the MRVM, though not yet strong evidence. However, this evidence is too weak to claim a detection of a global magnetospheric twist. Critically, the marginalized posterior distribution of the twist parameter $\lambda$ peaks at a low value and exhibits a broad, decaying tail toward the prior boundary, rather than accumulating at the boundary itself. This indicates that the data themselves prefer a negligible twist, and the inference is driven by the likelihood rather than by the choice of prior range. The posterior of $\lambda$ also remains strongly degenerate with the viewing geometry, particularly with $\chi$ and $\zeta$. Therefore, the result should be regarded as a possible hint of non-dipolar or twisted effects, but not as a statistically significant requirement for a twisted magnetosphere.

For 1E~1547.0$-$5408, the case for a twist is even weaker. This is consistent with the CRVM-only analysis of \cite{Taverna2026arXiv}, who found that the phase-dependent X-ray PA swing can be described by a dipole-based RVM. Under flat priors, the CRVM and MRVM have nearly identical maximum likelihoods, while the AIC and BIC favor the simpler CRVM. The Bayes factor is only ${\rm BF}=1.29$, which corresponds to negligible evidence according to\,\citet{Kass1995JASA}. The posterior predictive distributions of the two models are also nearly indistinguishable over the observed phase range. The posterior of $\lambda$ again peaks near zero, with a broad distribution that does not push against the prior boundary. Thus, the X-ray PA curve of 1E~1547.0$-$5408 does not require the additional twist correction included in the MRVM.

Although the data do not require a non-zero twist, the posterior distribution of $\lambda$ can be translated into an approximate upper limit on the global twist amplitude. In the self-similar twisted-dipole parameterization of\,\cite{Tong2021MNRAS}, $n=1-16\lambda^2/35$, and the maximum twist angle is approximately $\Delta\phi_{\max}\simeq2\lambda$ in the small-twist limit. For 1E~2259+586, the 84th percentile gives $\lambda<1.26$, corresponding to $n>0.27$ and $\Delta\phi_{\max}<2.52~{\rm rad}$. For 1E~1547.0$-$5408, the corresponding limits are $\lambda<1.30$ for the flat-prior fit and $\lambda<1.26$ for the radio-informed fit, corresponding to $n>0.23$ and $n>0.27$, respectively. These limits are indicative rather than stringent, because the 95\% upper limits approach the imposed prior boundary, $\lambda<1.48$. The present data therefore provide only weak, prior-limited constraints on the global twist amplitude. They do not require a twisted magnetosphere, but they also do not exclude moderately twisted configurations such as the illustrative $n\simeq0.5$ case discussed by\,\cite{Tong2021MNRAS}.

This conclusion should not be interpreted as evidence that magnetar magnetospheres are strictly dipolar. Magnetars are expected to host magnetospheric currents, twisted field lines, resonant scattering regions, and possibly multipolar surface fields\,\citep{Thompson2002ApJ,Beloborodov2009ApJ,Kaspi2017ARAA}. Our result is more limited: the current phase-resolved IXPE PA data are consistent with a dipole-dominated large-scale geometry. Any twist-related correction, if present, is either too small to isolate with the present data, strongly degenerate with the viewing geometry, or not fully captured by the first-order globally twisted MRVM adopted here.

It is also useful to distinguish the magnetospheric-twist interpretation considered here from the free-precession scenario proposed for the radio magnetar XTE~J1810$-$197. High-cadence radio polarimetric observations of XTE~J1810$-$197 after its 2018 outburst showed rapid and systematic changes in the PA curve, including reversals in the sign of the PA gradient, which have been interpreted as evidence for a freely precessing magnetar with a time-dependent viewing geometry\,\citep{Desvignes2024NatAs}. In that case, the main effect is a temporal variation of the effective RVM geometry, such as the impact angle and the line-of-sight cut through the emission beam. By contrast, the MRVM used in this work assumes a fixed viewing geometry for each data set and tests whether the PA curve at a given epoch requires an additional twist-induced correction. The two effects are therefore physically distinct: free precession probes the time-dependent orientation and deformation of the neutron star, whereas the MRVM parameter $\lambda$ probes a phenomenological global twist of the magnetosphere. Multi-epoch X-ray polarimetry would be required to separate a time-dependent viewing geometry from magnetospheric untwisting.

\subsection{Implications for the X-ray Polarization Diversity of Magnetars}
\label{subsec:discussion_diversity}

The two sources show markedly different X-ray polarization properties. 1E~2259+586 has a relatively low phase-averaged polarization degree, whereas 1E~1547.0$-$5408 is highly polarized\,\citep{Heyl2024MNRAS,Stewart2025arXiv}. Our results suggest that viewing geometry contributes to this difference, but cannot be the only factor.

For 1E~2259+586, both the CRVM and MRVM favor a relatively small impact angle, implying that the line of sight passes close to the magnetic axis during the spin cycle. Such a geometry can naturally produce a smooth PA swing, but it does not by itself explain the modest polarization degree. The low PD may instead reflect partial cancellation among multiple emission regions, magnetospheric scattering, or a mixture of ordinary and extraordinary polarization modes\,\citep{Taverna2020MNRAS}.

For 1E~1547.0$-$5408, the flat-prior X-ray fit favors a small magnetic inclination and a much larger viewing angle, implying a larger impact angle than in 1E~2259+586. The high PD may therefore be associated with a more coherent projected magnetic-field orientation or with the dominance of one polarization mode. However, the broad posterior on $\zeta$ shows that the detailed viewing geometry is still not uniquely determined by the PA curve alone. This is consistent with the well-known degeneracy of the RVM, where the PA curve primarily constrains the impact angle $\beta$ rather than the individual angles $\chi$ and $\zeta$ (see Section~\ref{sec:1547_results}).

The comparison of the two sources indicates that magnetar X-ray polarization diversity cannot be reduced to a single geometrical parameter. The observed PA and PD depend on the viewing geometry, but also on the surface temperature distribution, magnetic-field topology, vacuum birefringence, resonant scattering, and possible phase-dependent mode mixing\,\citep{Fern2007ApJ,Taverna2020MNRAS,Taverna2022Sci}. The geometrical modeling presented here provides a useful baseline, but a complete interpretation requires joint modeling of the phase-resolved flux, PD, and PA.

\subsection{X-ray and Radio RVM Geometries of 1E~1547.0$-$5408}
\label{subsec:discussion_radio_xray}

As discussed in Section~\ref{subsec:comparison_previous_1547}, our flat-prior X-ray geometry is closer to the X-ray RVM solutions of \cite{Taverna2026arXiv} and \cite{Stewart2025arXiv} than to the nearly aligned radio RVM geometry. The radio observations of 1E~1547.0$-$5408 favor a nearly aligned geometry, with $\chi_{\rm r}\simeq3.4^\circ$ and $\zeta_{\rm r}\simeq7.5^\circ$\,\citep{Camilo2008ApJ,Stewart2025arXiv}. When these constraints are imposed as informative priors, the X-ray posterior distributions are pulled toward the same nearly aligned configuration. The resulting fits remain statistically acceptable, indicating that the radio geometry is not ruled out by the X-ray data. However, the fit quality is poorer than in the flat-prior case, with $\chi^2_{\min}$ increasing from 3.58 to 6.19 for the CRVM and from 3.58 to 5.86 for the MRVM.

This behavior suggests that the radio and X-ray constraints are compatible, but not identical. The X-ray-only fit prefers a small magnetic inclination and a larger viewing angle, whereas the radio-informed fit is largely driven by the external prior. This does not necessarily imply different global magnetic geometries. Instead, the inferred RVM parameters should be understood as effective geometrical quantities that may depend on waveband, emission altitude, propagation effects, and the adopted polarization model\,\citep{Blaskiewicz1991ApJ,Everett2001ApJ,Johnston2019MNRAS}.

Several effects may contribute to this apparent difference. The radio emission is likely produced along open magnetospheric field lines, while the X-ray polarization may originate from surface thermal emission and be modified by vacuum birefringence and magnetospheric scattering\,\citep{Taverna2020MNRAS}. The radio PA curve may also be affected by orthogonal polarization modes or plasma propagation effects\,\citep{Lyne1988MNRAS,Everett2001ApJ}. A more rigorous comparison would require a joint radio--X-ray fit at the Stokes-parameter level, allowing for waveband-dependent emission heights, PA offsets, mode mixing, and different emission-region morphologies.

\subsection{Limitations of the Present Modeling}
\label{subsec:discussion_limitations}

Several limitations should be noted. First, our likelihood is based on published phase-resolved PA values and their uncertainties. This is adequate for a first geometrical comparison, but it is less fundamental than fitting the normalized Stokes parameters directly. In particular, when the polarization degree is low or comparable to the minimum detectable polarization, the PA uncertainty can become non-Gaussian\,\citep{Naghizadeh1993AA,Everett2001ApJ}. For 1E~2259+586, we reduce this issue by using only phase bins with PD above $\mathrm{MDP}_{99}$, retaining seven phase bins (see Section~\ref{sec:data}), but a future Stokes-level analysis would be more robust.

Second, both the CRVM and MRVM are purely geometrical models. They do not describe the phase-resolved flux, polarization degree, energy dependence, or radiative transfer in the magnetized atmosphere and magnetosphere. The inferred parameters should therefore be interpreted as effective geometrical constraints, not as a complete physical description of the source. In particular, the MRVM twist parameter $\lambda$ should not be over-interpreted as a unique measurement of the magnetospheric current structure.

Third, the posterior distributions show significant degeneracies, especially for 1E~1547.0$-$5408, where $\zeta$ remains broadly constrained under flat priors. Such degeneracies are expected when only the PA curve is fitted. Including phase-resolved PD and flux profiles, or jointly modeling multiple energy bands, would help break these degeneracies.

Finally, the present analysis is based on single-epoch X-ray polarimetric observations. Magnetars are variable objects, and their magnetospheric configurations may evolve with time, particularly during or after outbursts\,\citep{Beloborodov2009ApJ,Kaspi2017ARAA}. Multi-epoch IXPE observations, future X-ray polarimetric missions, and direct Stokes-level modeling will be essential for testing whether the inferred geometry and any possible twist-related correction remain stable over time.

\section{Summary}
\label{sec:summary}

We have presented a Bayesian analysis of IXPE phase-resolved polarization position angle (PA) data for two magnetars, 1E~2259+586 and 1E~1547.0$-$5408, comparing the CRVM with the MRVM that includes a first-order magnetospheric twist correction. Our key findings are as follows:

\begin{enumerate}
    \item The MRVM does not yield a statistically significant improvement over the CRVM for either source. For 1E~2259+586, the MRVM gives a lower $\chi^2_{\min}$ and lower AIC/BIC values, but the Bayesian evidence difference is only $\Delta\ln Z\simeq1.18 \quad ({\rm BF}\simeq3.25)$), which is too weak to claim a required global twist. For 1E~1547.0$-$5408, the CRVM and MRVM have nearly identical maximum likelihoods; the AIC and BIC favor the simpler CRVM, and $\Delta\ln Z\simeq0.26 \quad ({\rm BF}\simeq1.29)$) indicates no meaningful preference for the MRVM. This conclusion is consistent with the CRVM-only analysis of \cite{Taverna2026arXiv}, which also found that a dipole-based RVM can reproduce the X-ray PA swing.

    \item The inferred impact angle differs between the two sources and depends on the adopted model or prior. For 1E~2259+586, the CRVM gives $\beta\approx 11^\circ$, while the MRVM gives $\beta\approx 5^\circ$; both indicate a small-impact-angle geometry. For 1E~1547.0$-$5408 under flat priors, the CRVM favors a small magnetic inclination $\chi$ and a large viewing angle $\zeta$, giving $\beta\approx 65^\circ$. The radio-informed prior pulls the solution toward a nearly aligned configuration ($\chi\approx 3^\circ,\ \zeta\approx 9^\circ$), but the fit quality degrades, indicating that the X-ray data alone prefer a less-aligned effective geometry. Our flat-prior CRVM geometry for 1E~1547.0$-$5408 is broadly consistent with previous X-ray RVM results, while the radio-informed fit shows that the nearly aligned radio geometry of \cite{Stewart2025arXiv} is compatible with, but not independently preferred by, the X-ray PA data.

    \item The twist parameter $\lambda$ is not tightly constrained. The posterior distributions are broad in both sources, and the 95\% upper limits approach the imposed prior boundary, $\lambda<1.48$. Thus, the current data provide only weak, prior-limited constraints on the global twist amplitude. Thus, the present PA-only analysis does not identify a resolvable global twist as the origin of the radio--X-ray geometrical difference.
\end{enumerate}

Our study demonstrates that phase-resolved PA curves primarily constrain combinations of $\chi$ and $\zeta$, while degeneracies between the two angles persist. The observed diversity in polarization degree suggests that emission physics, in addition to geometry, plays a crucial role. Future higher-sensitivity observations with missions such as \textit{eXTP}, combined with full Stokes-parameter-level modeling, will be essential for breaking these degeneracies and probing magnetospheric physics in detail \citep{Zhang2025SCPMA,Ge2025SCPMA}.

\section*{Acknowledgements}

We would like to thank Mingyu Ge of the Institute of High Energy Physics, Chinese Academy of Sciences, for valuable suggestions that substantially improved this work. This research was supported by the National Key Research and Development Program of China (2022YFC2205202), the Major Science and Technology Special Project of Xinjiang Uygur Autonomous Region (2022A03013-1), the National Natural Science Foundation of China (12041304, 12288102, 12373114, 12003009), the Natural Science Foundation of Xinjiang Uygur Autonomous Region (2022D01A155), and the Tianshan Talents Program (2023TSYCTD0013).

\appendix
\restartappendixnumbering
\section{Posterior Distributions}
\label{app:posterior}

In this appendix, we present the posterior distributions of the geometrical parameters for the CRVM and MRVM fits. These corner plots are used to illustrate the parameter degeneracies discussed in Section~\ref{sec:results}.

\begin{figure*}[!htbp]
\centering
\gridline{
\fig{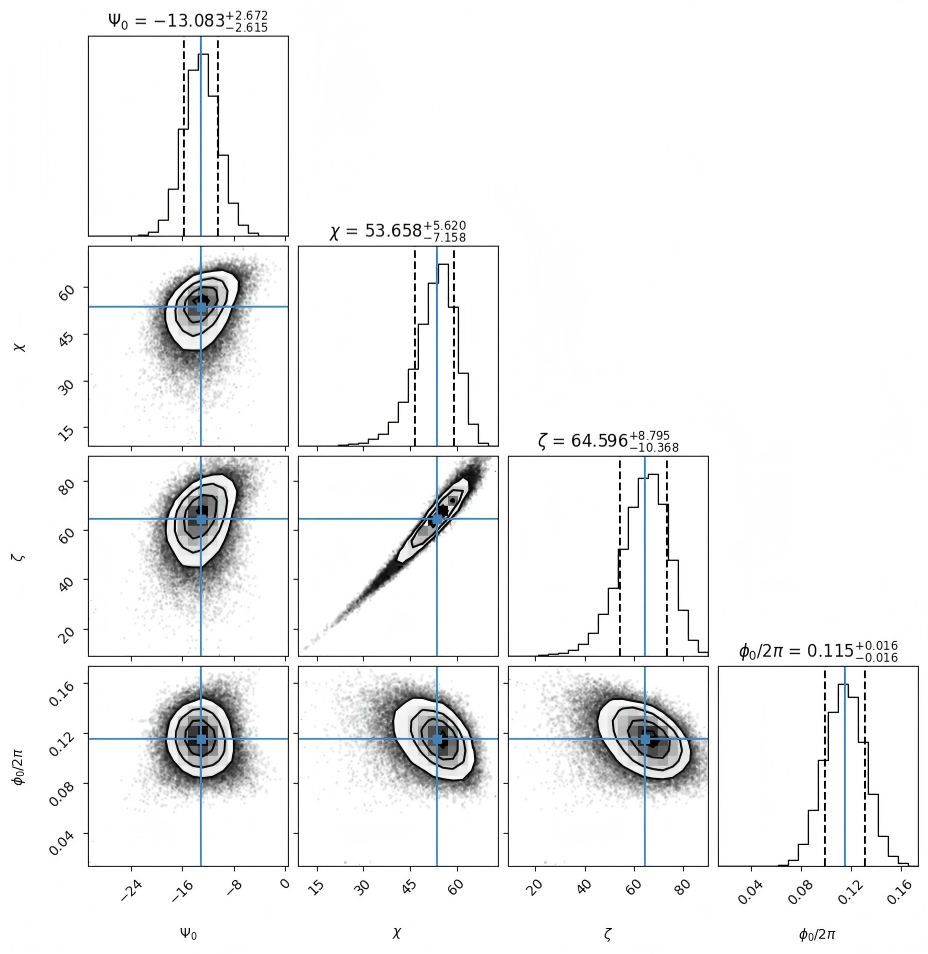}{0.48\textwidth}{(a) CRVM}
\fig{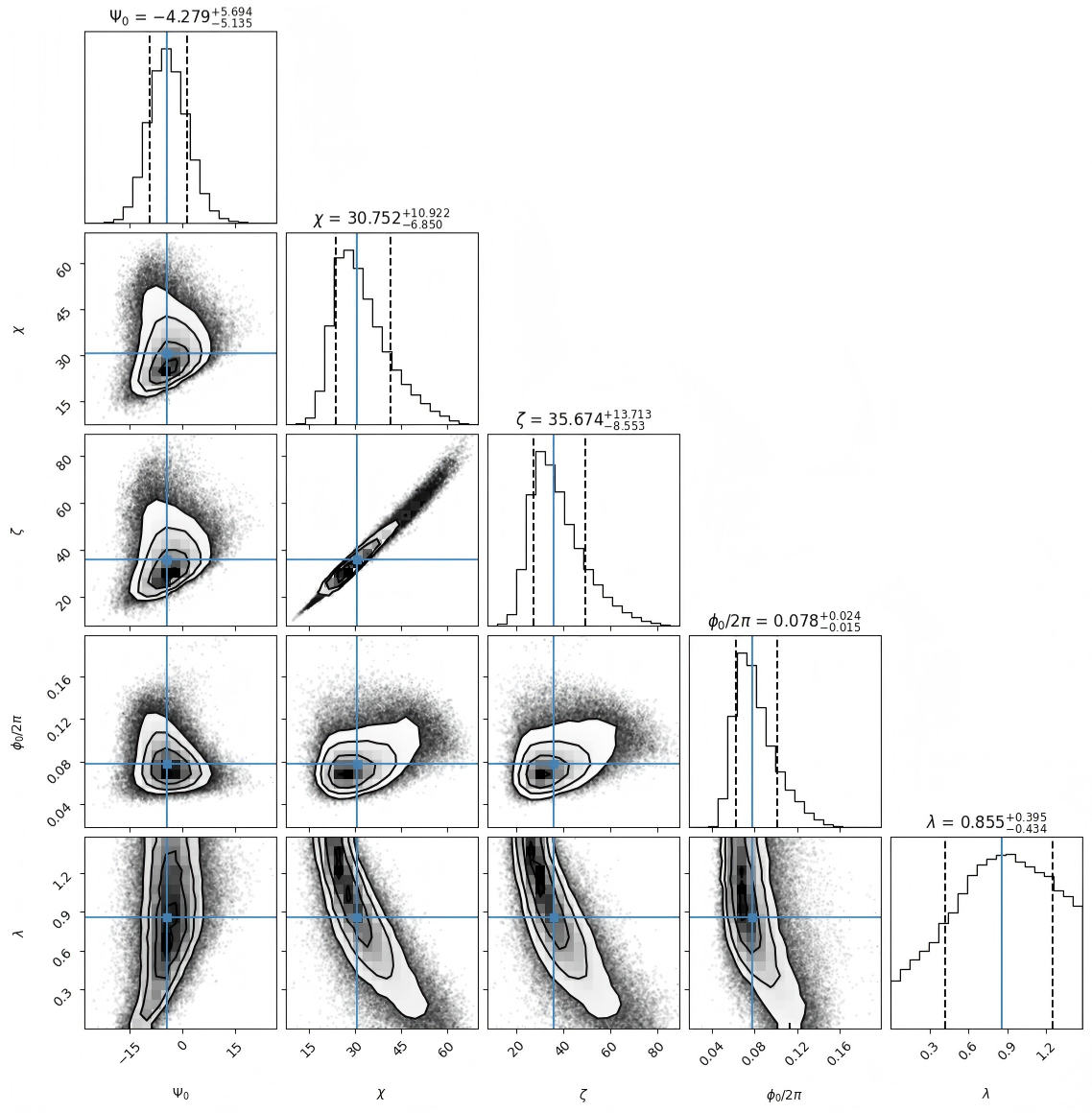}{0.48\textwidth}{(b) MRVM}
}
\caption{
Posterior corner plots for 1E~2259+586.
Panel (a) shows the CRVM result, and panel (b) shows the MRVM result.
Compared with the CRVM case, the MRVM posterior is broader because of the additional coupling between the geometrical parameters and the twist parameter $\lambda$.
}
\label{fig:app_corner_2259}
\end{figure*}

\begin{figure*}[!htbp]
\centering
\gridline{
\fig{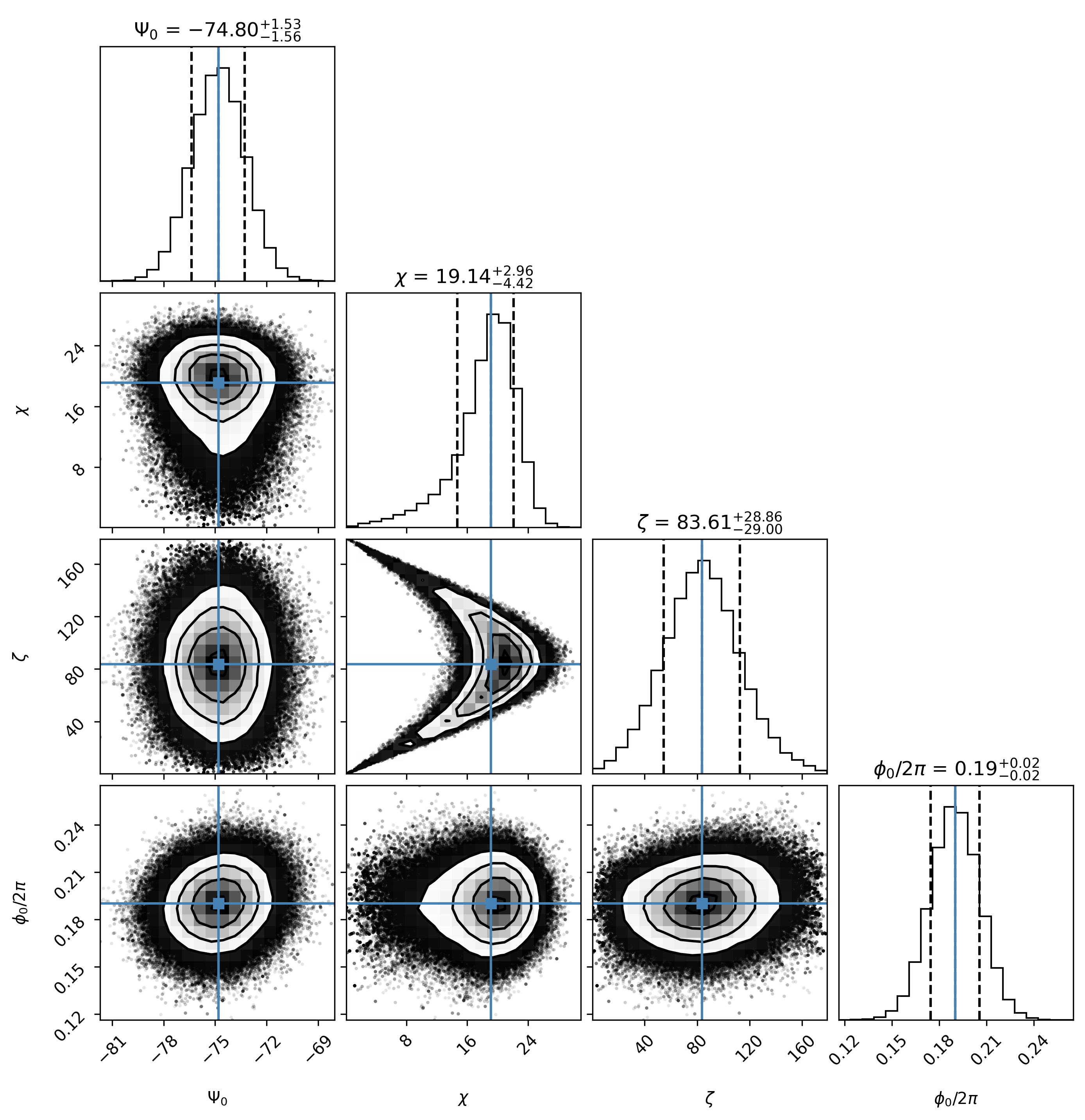}{0.48\textwidth}{(a) CRVM}
\fig{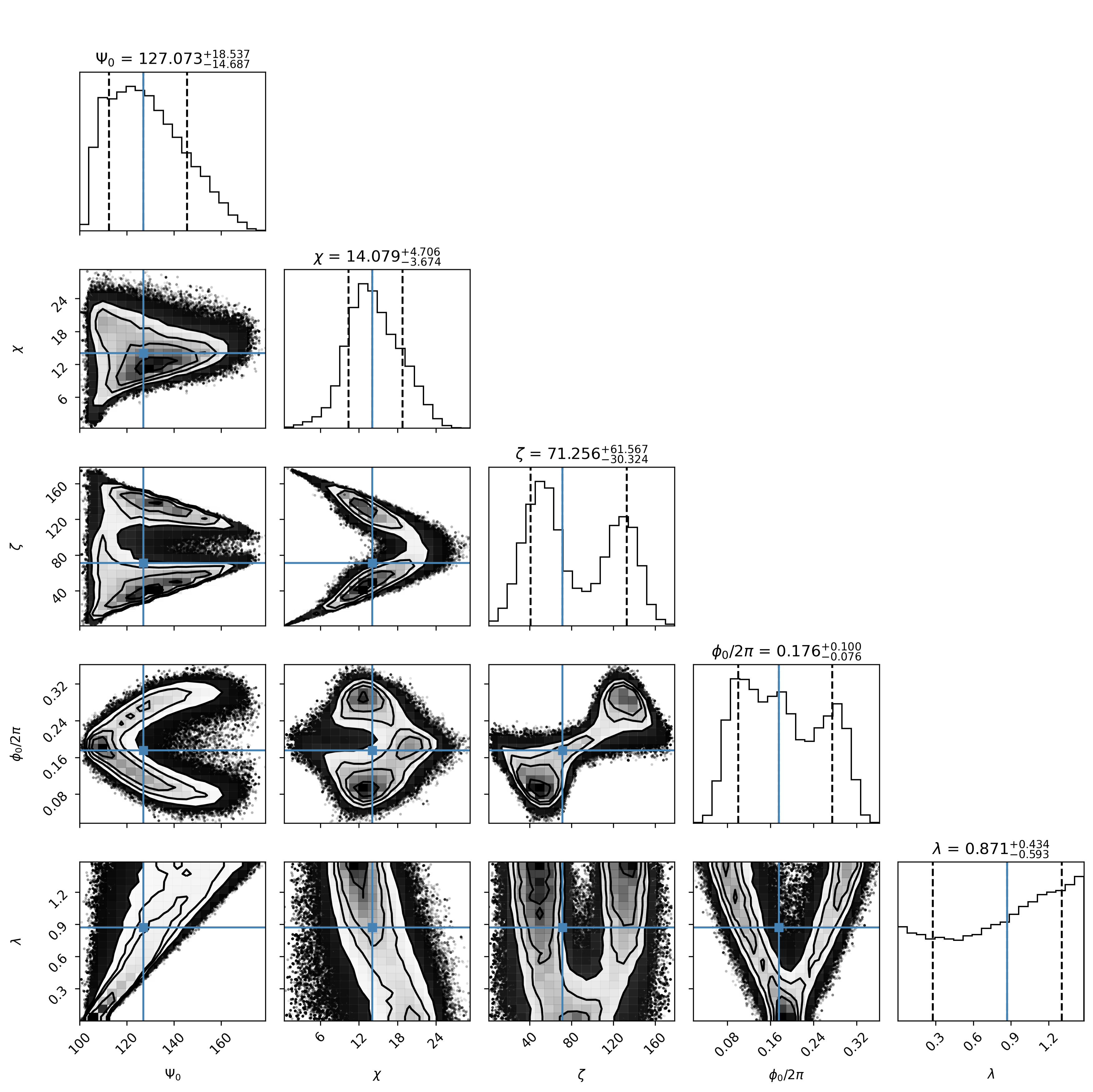}{0.48\textwidth}{(b) MRVM}
}
\caption{
Posterior corner plots for 1E~1547.0$-$5408 obtained with flat priors.
Panel (a) shows the CRVM result, and panel (b) shows the MRVM result.
The MRVM posterior is broader, especially in $\zeta$ and $\lambda$, reflecting the increased degeneracy introduced by the twist correction.
}
\label{fig:app_corner_1547_flat}
\end{figure*}

\begin{figure*}[!htbp]
\centering
\gridline{
\fig{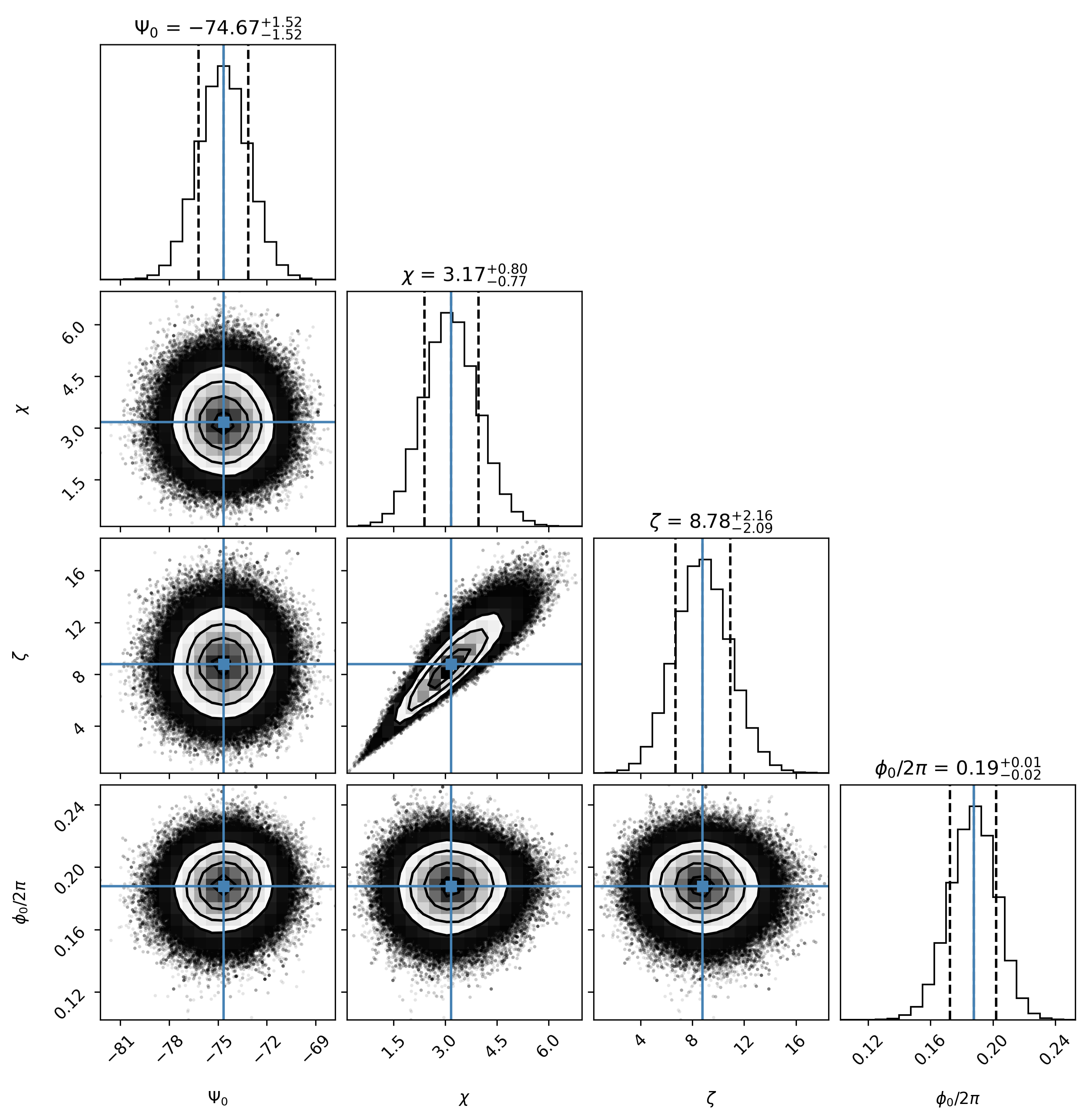}{0.48\textwidth}{(a) CRVM}
\fig{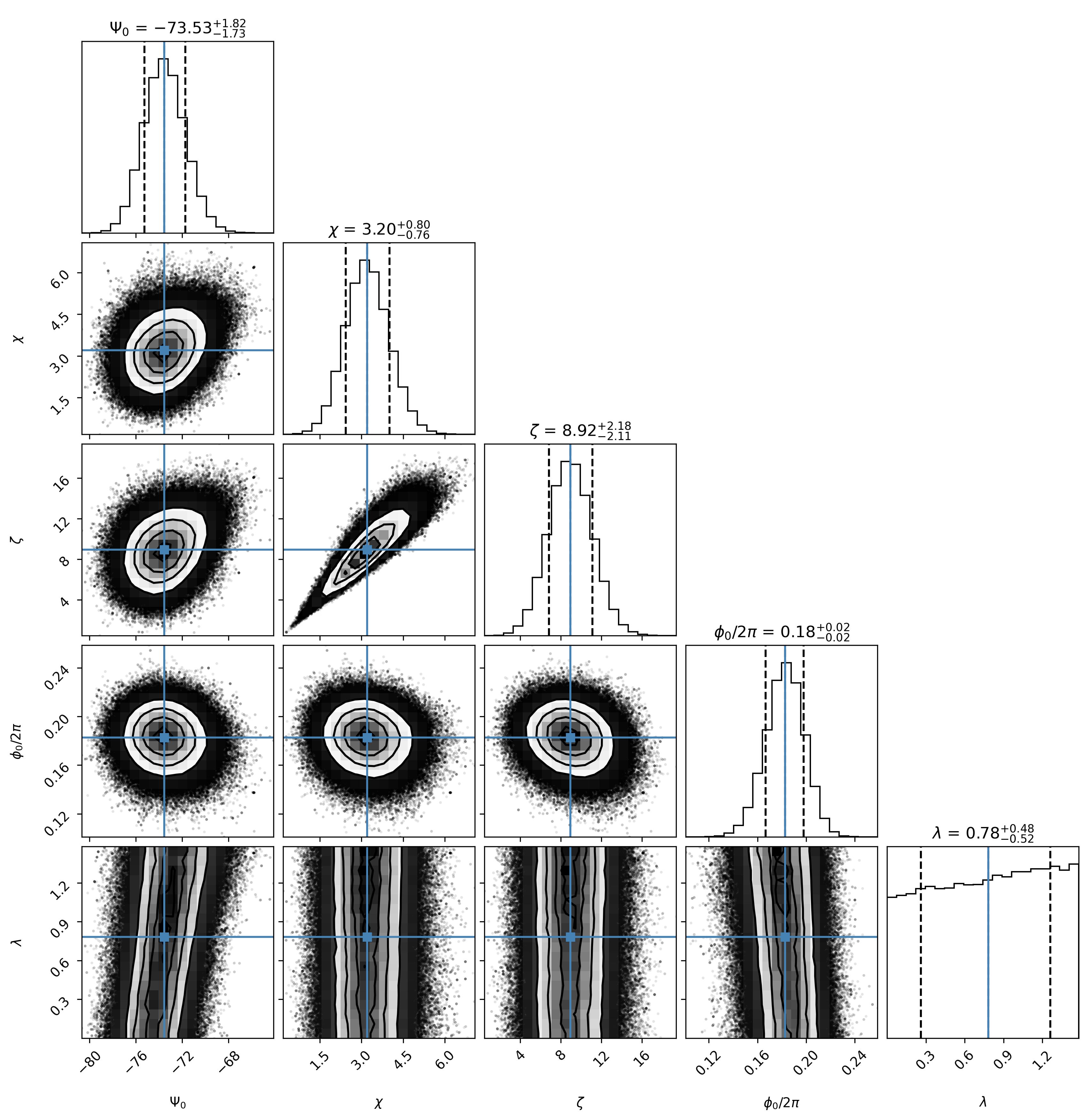}{0.48\textwidth}{(b) MRVM}
}
\caption{
Posterior corner plots for 1E~1547.0$-$5408 obtained with radio-informed priors.
Panel (a) shows the CRVM result, and panel (b) shows the MRVM result.
Under the radio-informed priors, both models are pulled toward a nearly aligned geometry, while the MRVM still retains a broad posterior distribution for the twist parameter $\lambda$.
}
\label{fig:app_corner_1547_radio}
\end{figure*}

\clearpage

\bibliography{ref}{}

@ARTICLE{Kaspi2017ARAA,
       author = {{Kaspi}, Victoria M. and {Beloborodov}, Andrei M.},
        title = "{Magnetars}",
      journal = {\araa},
         year = 2017,
        month = aug,
       volume = {55},
       number = {1},
        pages = {261-301},
          doi = {10.1146/annurev-astro-081915-023329},
archivePrefix = {arXiv},
       eprint = {1703.00068},
 primaryClass = {astro-ph.HE},
       adsurl = {https://ui.adsabs.harvard.edu/abs/2017ARA&A..55..261K}
}

@ARTICLE{Duncan1992ApJ,
       author = {{Duncan}, Robert C. and {Thompson}, Christopher},
        title = "{Formation of Very Strongly Magnetized Neutron Stars: Implications for Gamma-Ray Bursts}",
      journal = {\apjl},
         year = 1992,
        month = jun,
       volume = {392},
        pages = {L9},
          doi = {10.1086/186413},
       adsurl = {https://ui.adsabs.harvard.edu/abs/1992ApJ...392L...9D}
}

@ARTICLE{Olausen2014ApJS,
       author = {{Olausen}, S.~A. and {Kaspi}, V.~M.},
        title = "{The McGill Magnetar Catalog}",
      journal = {Astrophys. J. Suppl.},
         year = 2014,
        month = may,
       volume = {212},
       number = {1},
          eid = {6},
        pages = {6},
          doi = {10.1088/0067-0049/212/1/6},
archivePrefix = {arXiv},
       eprint = {1309.4167},
 primaryClass = {astro-ph.HE},
       adsurl = {https://ui.adsabs.harvard.edu/abs/2014ApJS..212....6O}
}

@ARTICLE{Turolla2015RPPh,
       author = {{Turolla}, R. and {Zane}, S. and {Watts}, A.~L.},
        title = "{Magnetars: the physics behind observations. A review}",
      journal = {Reports on Progress in Physics},
         year = 2015,
        month = nov,
       volume = {78},
       number = {11},
          eid = {116901},
        pages = {116901},
          doi = {10.1088/0034-4885/78/11/116901},
archivePrefix = {arXiv},
       eprint = {1507.02924},
 primaryClass = {astro-ph.HE},
       adsurl = {https://ui.adsabs.harvard.edu/abs/2015RPPh...78k6901T}
}

@ARTICLE{Thompson2002ApJ,
       author = {{Thompson}, C. and {Lyutikov}, M. and {Kulkarni}, S.~R.},
        title = "{Electrodynamics of Magnetars: Implications for the Persistent X-Ray Emission and Spin-down of the Soft Gamma Repeaters and Anomalous X-Ray Pulsars}",
      journal = {\apj},
         year = 2002,
        month = jul,
       volume = {574},
       number = {1},
        pages = {332-355},
          doi = {10.1086/340586},
archivePrefix = {arXiv},
       eprint = {astro-ph/0110677},
 primaryClass = {astro-ph},
       adsurl = {https://ui.adsabs.harvard.edu/abs/2002ApJ...574..332T}
}

@ARTICLE{Beloborodov2009ApJ,
       author = {{Beloborodov}, Andrei M.},
        title = "{Untwisting Magnetospheres of Neutron Stars}",
      journal = {\apj},
         year = 2009,
        month = sep,
       volume = {703},
       number = {1},
        pages = {1044-1060},
          doi = {10.1088/0004-637X/703/1/1044},
archivePrefix = {arXiv},
       eprint = {0812.4873},
 primaryClass = {astro-ph},
       adsurl = {https://ui.adsabs.harvard.edu/abs/2009ApJ...703.1044B}
}

@ARTICLE{LaiHo2003ApJ,
       author = {{Lai}, Dong and {Ho}, Wynn C.~G.},
        title = "{Transfer of Polarized Radiation in Strongly Magnetized Plasmas and Thermal Emission from Magnetars: Effect of Vacuum Polarization}",
      journal = {\apj},
         year = 2003,
        month = may,
       volume = {588},
       number = {2},
        pages = {962-974},
          doi = {10.1086/374334},
archivePrefix = {arXiv},
       eprint = {astro-ph/0211315},
 primaryClass = {astro-ph},
       adsurl = {https://ui.adsabs.harvard.edu/abs/2003ApJ...588..962L}
}

@ARTICLE{Taverna2014MNRAS,
       author = {{Taverna}, R. and {Muleri}, F. and {Turolla}, R. and {Soffitta}, P. and {Fabiani}, S. and {Nobili}, L.},
        title = "{Probing magnetar magnetosphere through X-ray polarization measurements}",
      journal = {\mnras},
         year = 2014,
        month = feb,
       volume = {438},
       number = {2},
        pages = {1686-1697},
          doi = {10.1093/mnras/stt2310},
archivePrefix = {arXiv},
       eprint = {1311.7500},
 primaryClass = {astro-ph.HE},
       adsurl = {https://ui.adsabs.harvard.edu/abs/2014MNRAS.438.1686T}
}

@ARTICLE{Taverna2022Sci,
       author = {{Taverna}, Roberto and {Turolla}, Roberto and {Muleri}, Fabio and {Heyl}, Jeremy and {Zane}, Silvia and {Baldini}, Luca and {Gonz{\'a}lez-Caniulef}, Denis and {Bachetti}, Matteo and {Rankin}, John and {Caiazzo}, Ilaria and {Di Lalla}, Niccol{\`o} and {Doroshenko}, Victor and {Errando}, Manel and {Gau}, Ephraim and {K{\i}rm{\i}z{\i}bayrak}, Demet and {Krawczynski}, Henric and {Negro}, Michela and {Ng}, Mason and {Omodei}, Nicola and {Possenti}, Andrea and {Tamagawa}, Toru and {Uchiyama}, Keisuke and {Weisskopf}, Martin C. and {Agudo}, Ivan and {Antonelli}, Lucio A. and {Baumgartner}, Wayne H. and {Bellazzini}, Ronaldo and {Bianchi}, Stefano and {Bongiorno}, Stephen D. and {Bonino}, Raffaella and {Brez}, Alessandro and {Bucciantini}, Niccol{\`o} and {Capitanio}, Fiamma and {Castellano}, Simone and {Cavazzuti}, Elisabetta and {Ciprini}, Stefano and {Costa}, Enrico and {De Rosa}, Alessandra and {Del Monte}, Ettore and {Di Gesu}, Laura and {Di Marco}, Alessandro and {Donnarumma}, Immacolata and {Dov{\v{c}}iak}, Michal and {Ehlert}, Steven R. and {Enoto}, Teruaki and {Evangelista}, Yuri and {Fabiani}, Sergio and {Ferrazzoli}, Riccardo and {Garcia}, Javier A. and {Gunji}, Shuichi and {Hayashida}, Kiyoshi and {Iwakiri}, Wataru and {Jorstad}, Svetlana G. and {Karas}, Vladimir and {Kitaguchi}, Takao and {Kolodziejczak}, Jeffery J. and {La Monaca}, Fabio and {Latronico}, Luca and {Liodakis}, Ioannis and {Maldera}, Simone and {Manfreda}, Alberto and {Marin}, Fr{\'e}d{\'e}ric and {Marinucci}, Andrea and {Marscher}, Alan P. and {Marshall}, Herman L. and {Matt}, Giorgio and {Mitsuishi}, Ikuyuki and {Mizuno}, Tsunefumi and {Ng}, Stephen C.-Y. and {O{\textquoteright}Dell}, Stephen L. and {Oppedisano}, Chiara and {Papitto}, Alessandro and {Pavlov}, George G. and {Peirson}, Abel L. and {Perri}, Matteo and {Pesce-Rollins}, Melissa and {Pilia}, Maura and {Poutanen}, Juri and {Puccetti}, Simonetta and {Ramsey}, Brian D. and {Ratheesh}, Ajay and {Romani}, Roger W. and {Sgr{\`o}}, Carmelo and {Slane}, Patrick and {Soffitta}, Paolo and {Spandre}, Gloria and {Tavecchio}, Fabrizio and {Tawara}, Yuzuru and {Tennant}, Allyn F. and {Thomas}, Nicholas E. and {Tombesi}, Francesco and {Trois}, Alessio and {Tsygankov}, Sergey S. and {Vink}, Jacco and {Wu}, Kinwah and {Xie}, Fei},
        title = "{Polarized x-rays from a magnetar}",
      journal = {Science},
         year = 2022,
        month = nov,
       volume = {378},
       number = {6620},
        pages = {646-650},
          doi = {10.1126/science.add0080},
archivePrefix = {arXiv},
       eprint = {2205.08898},
 primaryClass = {astro-ph.HE},
       adsurl = {https://ui.adsabs.harvard.edu/abs/2022Sci...378..646T}
}

@ARTICLE{Radhakrishnan1969ApL,
       author = {{Radhakrishnan}, V. and {Cooke}, D.~J.},
        title = "{Magnetic Poles and the Polarization Structure of Pulsar Radiation}",
      journal = {\aplett},
         year = 1969,
        month = jan,
       volume = {3},
        pages = {225},
       adsurl = {https://ui.adsabs.harvard.edu/abs/1969ApL.....3..225R}
}

@ARTICLE{Yuen2026MNRAS,
       author = {{Yuen}, R.},
        title = "{Explore ultra-long period radio pulsars and transients with pulsar emission geometry}",
      journal = {\mnras},
         year = 2026,
        month = apr,
       volume = {547},
       number = {3},
          eid = {stag395},
        pages = {stag395},
          doi = {10.1093/mnras/stag395},
       adsurl = {https://ui.adsabs.harvard.edu/abs/2026MNRAS.547ag395Y}
}

@ARTICLE{Yuen2024ApJ,
       author = {{Yuen}, R.},
        title = "{The Relationships between Emission Geometry and Jitter Noise in Millisecond Pulsars}",
      journal = {\apj},
         year = 2024,
        month = may,
       volume = {966},
       number = {1},
          eid = {34},
        pages = {34},
          doi = {10.3847/1538-4357/ad2e05},
       adsurl = {https://ui.adsabs.harvard.edu/abs/2024ApJ...966...34Y}
}

@ARTICLE{Wen2022ApJ,
       author = {{Wen}, Z.~G. and {Yuan}, J.~P. and {Wang}, N. and {Li}, D. and {Chen}, J.~L. and {Wang}, P. and {Wu}, Q.~D. and {Yan}, W.~M. and {Yuen}, R. and {Wang}, Z. and {Tedila}, H.~M. and {Wang}, H.~G. and {Zhu}, W.~W. and {Niu}, J.~R. and {Miao}, C.~C. and {Xue}, M.~Y. and {Duan}, X.~F. and {Xiang}, B.~B. and {He}, D.~L.},
        title = "{A Single-pulse Study of the Subpulse Drifter PSR J1631+1252 Discovered at FAST}",
      journal = {\apj},
         year = 2022,
        month = apr,
       volume = {929},
       number = {1},
          eid = {71},
        pages = {71},
          doi = {10.3847/1538-4357/ac5d5d},
       adsurl = {https://ui.adsabs.harvard.edu/abs/2022ApJ...929...71W}
}

@ARTICLE{Blaskiewicz1991ApJ,
       author = {{Blaskiewicz}, M. and {Cordes}, J.~M. and {Wasserman}, I.},
        title = "{A Relativistic Model of Pulsar Polarization}",
      journal = {\apj},
         year = 1991,
        month = apr,
       volume = {370},
        pages = {643},
          doi = {10.1086/169850},
       adsurl = {https://ui.adsabs.harvard.edu/abs/1991ApJ...370..643B}
}

@ARTICLE{Weisskopf2022JATIS,
       author = {{Weisskopf}, Martin C. and {Soffitta}, Paolo and {Baldini}, Luca and {Ramsey}, Brian D. and {O'Dell}, Stephen L. and {Romani}, Roger W. and {Matt}, Giorgio and {Deininger}, William D. and {Baumgartner}, Wayne H. and {Bellazzini}, Ronaldo and {Costa}, Enrico and {Kolodziejczak}, Jeffery J. and {Latronico}, Luca and {Marshall}, Herman L. and {Muleri}, Fabio and {Bongiorno}, Stephen D. and {Tennant}, Allyn and {Bucciantini}, Niccolo and {Dovciak}, Michal and {Marin}, Frederic and {Marscher}, Alan and {Poutanen}, Juri and {Slane}, Pat and {Turolla}, Roberto and {Kalinowski}, William and {Di Marco}, Alessandro and {Fabiani}, Sergio and {Minuti}, Massimo and {La Monaca}, Fabio and {Pinchera}, Michele and {Rankin}, John and {Sgro'}, Carmelo and {Trois}, Alessio and {Xie}, Fei and {Alexander}, Cheryl and {Allen}, D. Zachery and {Amici}, Fabrizio and {Andersen}, Jason and {Antonelli}, Angelo and {Antoniak}, Spencer and {Attin{\`a}}, Primo and {Barbanera}, Mattia and {Bachetti}, Matteo and {Baggett}, Randy M. and {Bladt}, Jeff and {Brez}, Alessandro and {Bonino}, Raffaella and {Boree}, Christopher and {Borotto}, Fabio and {Breeding}, Shawn and {Brienza}, Daniele and {Bygott}, H. Kyle and {Caporale}, Ciro and {Cardelli}, Claudia and {Carpentiero}, Rita and {Castellano}, Simone and {Castronuovo}, Marco and {Cavalli}, Luca and {Cavazzuti}, Elisabetta and {Ceccanti}, Marco and {Centrone}, Mauro and {Citraro}, Saverio and {D'Amico}, Fabio and {D'Alba}, Elisa and {Di Gesu}, Laura and {Del Monte}, Ettore and {Dietz}, Kurtis L. and {Di Lalla}, Niccolo' and {Persio}, Giuseppe Di and {Dolan}, David and {Donnarumma}, Immacolata and {Evangelista}, Yuri and {Ferrant}, Kevin and {Ferrazzoli}, Riccardo and {Ferrie}, MacKenzie and {Footdale}, Joseph and {Forsyth}, Brent and {Foster}, Michelle and {Garelick}, Benjamin and {Gunji}, Shuichi and {Gurnee}, Eli and {Head}, Michael and {Hibbard}, Grant and {Johnson}, Samantha and {Kelly}, Erik and {Kilaru}, Kiranmayee and {Lefevre}, Carlo and {Roy}, Shelley Le and {Loffredo}, Pasqualino and {Lorenzi}, Paolo and {Lucchesi}, Leonardo and {Maddox}, Tyler and {Magazzu}, Guido and {Maldera}, Simone and {Manfreda}, Alberto and {Mangraviti}, Elio and {Marengo}, Marco and {Marrocchesi}, Alessandra and {Massaro}, Francesco and {Mauger}, David and {McCracken}, Jeffrey and {McEachen}, Michael and {Mize}, Rondal and {Mereu}, Paolo and {Mitchell}, Scott and {Mitsuishi}, Ikuyuki and {Morbidini}, Alfredo and {Mosti}, Federico and {Nasimi}, Hikmat and {Negri}, Barbara and {Negro}, Michela and {Nguyen}, Toan and {Nitschke}, Isaac and {Nuti}, Alessio and {Onizuka}, Mitch and {Oppedisano}, Chiara and {Orsini}, Leonardo and {Osborne}, Darren and {Pacheco}, Richard and {Paggi}, Alessandro and {Painter}, Will and {Pavelitz}, Steven D. and {Pentz}, Christina and {Piazzolla}, Raffaele and {Perri}, Matteo and {Pesce-Rollins}, Melissa and {Peterson}, Colin and {Pilia}, Maura and {Profeti}, Alessandro and {Puccetti}, Simonetta and {Ranganathan}, Jaganathan and {Ratheesh}, Ajay and {Reedy}, Lee and {Root}, Noah and {Rubini}, Alda and {Ruswick}, Stephanie and {Sanchez}, Javier and {Sarra}, Paolo and {Santoli}, Francesco and {Scalise}, Emanuele and {Sciortino}, Andrea and {Schroeder}, Christopher and {Seek}, Tim and {Sosdian}, Kalie and {Spandre}, Gloria and {Speegle}, Chet O. and {Tamagawa}, Toru and {Tardiola}, Marcello and {Tobia}, Antonino and {Thomas}, Nicholas E. and {Valerie}, Robert and {Vimercati}, Marco and {Walden}, Amy L. and {Weddendorf}, Bruce and {Wedmore}, Jeffrey and {Welch}, David and {Zanetti}, Davide and {Zanetti}, Francesco},
        title = "{The Imaging X-Ray Polarimetry Explorer (IXPE): Pre-Launch}",
      journal = {Journal of Astronomical Telescopes, Instruments, and Systems},
         year = 2022,
        month = apr,
       volume = {8},
       number = {2},
          eid = {026002},
        pages = {026002},
          doi = {10.1117/1.JATIS.8.2.026002},
archivePrefix = {arXiv},
       eprint = {2112.01269},
 primaryClass = {astro-ph.IM},
       adsurl = {https://ui.adsabs.harvard.edu/abs/2022JATIS...8b6002W}
}

@ARTICLE{Heyl2024MNRAS,
       author = {{Heyl}, Jeremy and {Taverna}, Roberto and {Turolla}, Roberto and {Israel}, Gian Luca and {Ng}, Mason and {K{\i}rm{\i}z{\i}bayrak}, Demet and {Gonz{\'a}lez-Caniulef}, Denis and {Caiazzo}, Ilaria and {Zane}, Silvia and {Ehlert}, Steven R. and {Negro}, Michela and {Agudo}, Iv{\'a}n and {Antonelli}, Lucio Angelo and {Bachetti}, Matteo and {Baldini}, Luca and {Baumgartner}, Wayne H. and {Bellazzini}, Ronaldo and {Bianchi}, Stefano and {Bongiorno}, Stephen D. and {Bonino}, Raffaella and {Brez}, Alessandro and {Bucciantini}, Niccol{\`o} and {Capitanio}, Fiamma and {Castellano}, Simone and {Cavazzuti}, Elisabetta and {Chen}, Chien-Ting and {Ciprini}, Stefano and {Costa}, Enrico and {De Rosa}, Alessandra and {Del Monte}, Ettore and {Di Gesu}, Laura and {Di Lalla}, Niccol{\`o} and {Di Marco}, Alessandro and {Donnarumma}, Immacolata and {Doroshenko}, Victor and {Dov{\v{c}}iak}, Michal and {Enoto}, Teruaki and {Evangelista}, Yuri and {Fabiani}, Sergio and {Ferrazzoli}, Riccardo and {Garcia}, Javier A. and {Gunji}, Shuichi and {Hayashida}, Kiyoshi and {Iwakiri}, Wataru and {Jorstad}, Svetlana G. and {Kaaret}, Philip and {Karas}, Vladimir and {Kislat}, Fabian and {Kitaguchi}, Takao and {Kolodziejczak}, Jeffery J. and {Krawczynski}, Henric and {La Monaca}, Fabio and {Latronico}, Luca and {Liodakis}, Ioannis and {Maldera}, Simone and {Manfreda}, Alberto and {Marin}, Fr{\'e}d{\'e}ric and {Marinucci}, Andrea and {Marscher}, Alan P. and {Marshall}, Herman L. and {Massaro}, Francesco and {Matt}, Giorgio and {Mitsuishi}, Ikuyuki and {Mizuno}, Tsunefumi and {Muleri}, Fabio and {Ng}, C.-Y. and {O'Dell}, Stephen L. and {Omodei}, Nicola and {Oppedisano}, Chiara and {Papitto}, Alessandro and {Pavlov}, George G. and {Peirson}, Abel Lawrence and {Perri}, Matteo and {Pesce-Rollins}, Melissa and {Petrucci}, Pierre-Olivier and {Pilia}, Maura and {Possenti}, Andrea and {Poutanen}, Juri and {Puccetti}, Simonetta and {Ramsey}, Brian D. and {Rankin}, John and {Ratheesh}, Ajay and {Roberts}, Oliver J. and {Romani}, Roger W. and {Sgr{\`o}}, Carmelo and {Slane}, Patrick and {Soffitta}, Paolo and {Spandre}, Gloria and {Swartz}, Douglas A. and {Tamagawa}, Toru and {Tavecchio}, Fabrizio and {Tawara}, Yuzuru and {Tennant}, Allyn F. and {Thomas}, Nicholas E. and {Tombesi}, Francesco and {Trois}, Alessio and {Tsygankov}, Sergey S. and {Vink}, Jacco and {Weisskopf}, Martin C. and {Wu}, Kinwah and {Xie}, Fei},
        title = "{The detection of polarized X-ray emission from the magnetar 1E 2259+586}",
      journal = {\mnras},
         year = 2024,
        month = feb,
       volume = {527},
       number = {4},
        pages = {12219-12231},
          doi = {10.1093/mnras/stad3680},
archivePrefix = {arXiv},
       eprint = {2311.03637},
 primaryClass = {astro-ph.HE},
       adsurl = {https://ui.adsabs.harvard.edu/abs/2024MNRAS.52712219H}
}

@ARTICLE{Camilo2008ApJ,
       author = {{Camilo}, F. and {Reynolds}, J. and {Johnston}, S. and {Halpern}, J.~P. and {Ransom}, S.~M.},
        title = "{The Magnetar 1E 1547.0-5408: Radio Spectrum, Polarimetry, and Timing}",
      journal = {\apj},
         year = 2008,
        month = may,
       volume = {679},
       number = {1},
        pages = {681-686},
          doi = {10.1086/587054},
archivePrefix = {arXiv},
       eprint = {0802.0494},
 primaryClass = {astro-ph},
       adsurl = {https://ui.adsabs.harvard.edu/abs/2008ApJ...679..681C}
}

@ARTICLE{Stewart2025arXiv,
       author = {{Stewart}, Rachael E. and {Dinh Thi}, Hoa and {Younes}, George and {Lower}, Marcus E. and {Baring}, Matthew G. and {Negro}, Michela and {Camilo}, Fernando and {Coley}, Joel B. and {Enoto}, Teruaki and {Harding}, Alice K. and {Ho}, Wynn C.~G. and {Hu}, Chin-Ping and {Kaaret}, Philip and {Scholz}, Paul and {Van Kooten}, Alex and {Wadiasingh}, Zorawar},
        title = "{Evidence of magnetospheric vacuum birefringence in the polarized X-rays of a radio magnetar}",
      journal = {arXiv e-prints},
         year = 2025,
        month = sep,
          eid = {arXiv:2509.19446},
        pages = {arXiv:2509.19446},
          doi = {10.48550/arXiv.2509.19446},
archivePrefix = {arXiv},
       eprint = {2509.19446},
 primaryClass = {astro-ph.HE},
       adsurl = {https://ui.adsabs.harvard.edu/abs/2025arXiv250919446S}
}

@ARTICLE{Taverna2026arXiv,
       author = {{Taverna}, Roberto and {Turolla}, Roberto and {Marra}, Lorenzo and {Kelly}, Ruth M.~E. and {Borghese}, Alice and {Israel}, Gian Luca and {Mereghetti}, Sandro and {Possenti}, Andrea and {Zane}, Silvia and {Rigoselli}, Michela},
        title = "{The Long Quest for Vacuum Birefringence in Magnetars: 1E 1547.0─5408 and the Elusive Smoking Gun}",
      journal = {\apj},
         year = 2026,
        month = may,
       volume = {1002},
       number = {1},
          eid = {102},
        pages = {102},
          doi = {10.3847/1538-4357/ae5c9d},
archivePrefix = {arXiv},
       eprint = {2601.15452},
 primaryClass = {astro-ph.HE},
       adsurl = {https://ui.adsabs.harvard.edu/abs/2026ApJ..1002..102T}
}

@ARTICLE{Gnedin1978SvAL,
  author  = {{Gnedin}, Yu. N. and {Pavlov}, G. G. and {Shibanov}, Yu. A.},
  title   = {The effect of vacuum birefringence in a magnetic field on polarization and directivity of radiation of X-ray pulsars},
  journal = {Soviet Astronomy Letters},
  year    = {1978},
  volume  = {4},
  pages   = {117--119}
}

@ARTICLE{Pavlov1979JETP,
       author = {{Pavlov}, G.~G. and {Shibanov}, Iu. A.},
        title = "{Influence of vacuum polarization by a magnetic field on the propagation of electromagnetic waves in plasmas}",
      journal = {Zhurnal Eksperimentalnoi i Teoreticheskoi Fiziki},
         year = 1979,
        month = may,
       volume = {76},
        pages = {1457-1473},
       adsurl = {https://ui.adsabs.harvard.edu/abs/1979ZhETF..76.1457P}
}

@ARTICLE{Heyl2000MNRAS,
       author = {{Heyl}, Jeremy S. and {Shaviv}, Nir J.},
        title = "{Polarization evolution in strong magnetic fields}",
      journal = {\mnras},
         year = 2000,
        month = jan,
       volume = {311},
       number = {3},
        pages = {555-564},
          doi = {10.1046/j.1365-8711.2000.03076.x},
archivePrefix = {arXiv},
       eprint = {astro-ph/9909339},
 primaryClass = {astro-ph},
       adsurl = {https://ui.adsabs.harvard.edu/abs/2000MNRAS.311..555H}
}

@ARTICLE{Heyl2002PRD,
       author = {{Heyl}, Jeremy S. and {Shaviv}, Nir J.},
        title = "{QED and the high polarization of the thermal radiation from neutron stars}",
      journal = {\prd},
         year = 2002,
        month = jul,
       volume = {66},
       number = {2},
          eid = {023002},
        pages = {023002},
          doi = {10.1103/PhysRevD.66.023002},
archivePrefix = {arXiv},
       eprint = {astro-ph/0203058},
 primaryClass = {astro-ph},
       adsurl = {https://ui.adsabs.harvard.edu/abs/2002PhRvD..66b3002H}
}

@ARTICLE{Taverna2015MNRAS,
       author = {{Taverna}, R. and {Turolla}, R. and {Gonzalez Caniulef}, D. and {Zane}, S. and {Muleri}, F. and {Soffitta}, P.},
        title = "{Polarization of neutron star surface emission: a systematic analysis}",
      journal = {\mnras},
         year = 2015,
        month = dec,
       volume = {454},
       number = {3},
        pages = {3254-3266},
          doi = {10.1093/mnras/stv2168},
archivePrefix = {arXiv},
       eprint = {1509.05023},
 primaryClass = {astro-ph.HE},
       adsurl = {https://ui.adsabs.harvard.edu/abs/2015MNRAS.454.3254T}
}

@ARTICLE{HeylCaiazzo2018Galaxies,
       author = {{Heyl}, Jeremy and {Caiazzo}, Ilaria},
        title = "{Strongly Magnetized Sources: QED and X-ray Polarization}",
      journal = {Galaxies},
         year = 2018,
        month = jul,
       volume = {6},
       number = {3},
          eid = {76},
        pages = {76},
          doi = {10.3390/galaxies6030076},
archivePrefix = {arXiv},
       eprint = {1802.00358},
 primaryClass = {astro-ph.HE},
       adsurl = {https://ui.adsabs.harvard.edu/abs/2018Galax...6...76H}
}

@ARTICLE{Poutanen2020AA,
       author = {{Poutanen}, Juri},
        title = "{Relativistic rotating vector model for X-ray millisecond pulsars}",
      journal = {\aap},
         year = 2020,
        month = sep,
       volume = {641},
          eid = {A166},
        pages = {A166},
          doi = {10.1051/0004-6361/202038689},
archivePrefix = {arXiv},
       eprint = {2006.10448},
 primaryClass = {astro-ph.HE},
       adsurl = {https://ui.adsabs.harvard.edu/abs/2020A&A...641A.166P}
}

@ARTICLE{GonzalezCaniulef2023MNRAS,
       author = {{Gonz{\'a}lez-Caniulef}, Denis and {Caiazzo}, Ilaria and {Heyl}, Jeremy},
        title = "{Unbinned likelihood analysis for X-ray polarization}",
      journal = {\mnras},
         year = 2023,
        month = mar,
       volume = {519},
       number = {4},
        pages = {5902-5912},
          doi = {10.1093/mnras/stad033},
archivePrefix = {arXiv},
       eprint = {2204.00140},
 primaryClass = {astro-ph.IM},
       adsurl = {https://ui.adsabs.harvard.edu/abs/2023MNRAS.519.5902G}
}

@ARTICLE{LaiHo2002ApJ,
       author = {{Lai}, Dong and {Ho}, Wynn C.~G.},
        title = "{Resonant Conversion of Photon Modes Due to Vacuum Polarization in a Magnetized Plasma: Implications for X-Ray Emission from Magnetars}",
      journal = {\apj},
         year = 2002,
        month = feb,
       volume = {566},
       number = {1},
        pages = {373-377},
          doi = {10.1086/338074},
archivePrefix = {arXiv},
       eprint = {astro-ph/0108127},
 primaryClass = {astro-ph},
       adsurl = {https://ui.adsabs.harvard.edu/abs/2002ApJ...566..373L}
}

@ARTICLE{Tong2021MNRAS,
       author = {{Tong}, H. and {Wang}, P.~F. and {Wang}, H.~G. and {Yan}, Z.},
        title = "{Rotating vector model for magnetars}",
      journal = {\mnras},
         year = 2021,
        month = mar,
       volume = {502},
       number = {1},
        pages = {1549-1556},
          doi = {10.1093/mnras/stab108},
archivePrefix = {arXiv},
       eprint = {2101.04504},
 primaryClass = {astro-ph.HE},
       adsurl = {https://ui.adsabs.harvard.edu/abs/2021MNRAS.502.1549T}
}

@ARTICLE{Storn1997JGOpt,
       author = {{Storn}, Rainer and {Price}, Kenneth},
        title = "{Differential Evolution - A Simple and Efficient Heuristic for global Optimization over Continuous Spaces}",
      journal = {Journal of Global Optimization},
         year = 1997,
        month = dec,
       volume = {11},
        pages = {341-359},
          doi = {10.1023/A:1008202821328},
       adsurl = {https://ui.adsabs.harvard.edu/abs/1997JGOpt..11..341S}
}

@ARTICLE{ForemanMackey2013PASP,
       author = {{Foreman-Mackey}, Daniel and {Hogg}, David W. and {Lang}, Dustin and {Goodman}, Jonathan},
        title = "{emcee: The MCMC Hammer}",
      journal = {\pasp},
         year = 2013,
        month = mar,
       volume = {125},
       number = {925},
        pages = {306},
          doi = {10.1086/670067},
archivePrefix = {arXiv},
       eprint = {1202.3665},
 primaryClass = {astro-ph.IM},
       adsurl = {https://ui.adsabs.harvard.edu/abs/2013PASP..125..306F}
}

@ARTICLE{Taverna2024Galax,
       author = {{Taverna}, Roberto and {Turolla}, Roberto},
        title = "{X-ray Polarization from Magnetar Sources}",
      journal = {Galaxies},
         year = 2024,
        month = feb,
       volume = {12},
       number = {1},
          eid = {6},
        pages = {6},
          doi = {10.3390/galaxies12010006},
archivePrefix = {arXiv},
       eprint = {2402.05622},
 primaryClass = {astro-ph.HE},
       adsurl = {https://ui.adsabs.harvard.edu/abs/2024Galax..12....6T}
}

@ARTICLE{Akaike1974ITAC,
       author = {{Akaike}, H.},
        title = "{A New Look at the Statistical Model Identification}",
      journal = {IEEE Transactions on Automatic Control},
         year = 1974,
        month = jan,
       volume = {19},
        pages = {716-723},
          doi = {10.1109/TAC.1974.1100705},
       adsurl = {https://ui.adsabs.harvard.edu/abs/1974ITAC...19..716A}
}

@ARTICLE{Schwarz1978AnSta,
       author = {{Schwarz}, Gideon},
        title = "{Estimating the Dimension of a Model}",
      journal = {Annals of Statistics},
         year = 1978,
        month = jul,
       volume = {6},
       number = {2},
        pages = {461-464},
       adsurl = {https://ui.adsabs.harvard.edu/abs/1978AnSta...6..461S}
}

@article{Kass1995JASA,
  author  = {{Kass}, R. E. and {Raftery}, A. E.},
  title   = {Bayes Factors},
  journal = {Journal of the American Statistical Association},
  year    = {1995},
  volume  = {90},
  number  = {430},
  pages   = {773--795},
  doi     = {10.1080/01621459.1995.10476572}
}

@ARTICLE{Gabry2017arXiv,
       author = {{Gabry}, Jonah and {Simpson}, Daniel and {Vehtari}, Aki and {Betancourt}, Michael and {Gelman}, Andrew},
        title = "{Visualization in Bayesian workflow}",
      journal = {arXiv e-prints},
         year = 2017,
        month = sep,
          eid = {arXiv:1709.01449},
        pages = {arXiv:1709.01449},
          doi = {10.48550/arXiv.1709.01449},
archivePrefix = {arXiv},
       eprint = {1709.01449},
 primaryClass = {stat.ME},
       adsurl = {https://ui.adsabs.harvard.edu/abs/2017arXiv170901449G}
}

@ARTICLE{Trotta2008ConPh,
       author = {{Trotta}, Roberto},
        title = "{Bayes in the sky: Bayesian inference and model selection in cosmology}",
      journal = {Contemporary Physics},
         year = 2008,
        month = mar,
       volume = {49},
       number = {2},
        pages = {71-104},
          doi = {10.1080/00107510802066753},
archivePrefix = {arXiv},
       eprint = {0803.4089},
 primaryClass = {astro-ph},
       adsurl = {https://ui.adsabs.harvard.edu/abs/2008ConPh..49...71T}
}

@ARTICLE{Speagle2020MNRAS,
       author = {{Speagle}, Joshua S.},
        title = "{DYNESTY: a dynamic nested sampling package for estimating Bayesian posteriors and evidences}",
      journal = {\mnras},
         year = 2020,
        month = apr,
       volume = {493},
       number = {3},
        pages = {3132-3158},
          doi = {10.1093/mnras/staa278},
archivePrefix = {arXiv},
       eprint = {1904.02180},
 primaryClass = {astro-ph.IM},
       adsurl = {https://ui.adsabs.harvard.edu/abs/2020MNRAS.493.3132S}
}

@ARTICLE{Zhang2025SCPMA,
       author = {{Zhang}, Shuang-Nan and {Santangelo}, Andrea and {Xu}, Yupeng and {Feng}, Hua and {Lu}, Fangjun and {Chen}, Yong and {Ge}, Mingyu and {Nandra}, Kirpal and {Wu}, Xin and {Feroci}, Marco and {Hernanz}, Margarita and {Liu}, Congzhan and {He}, Huilin and {Wang}, Yusa and {Jiang}, Weichun and {Cui}, Weiwei and {Yang}, Yanji and {Wang}, Juan and {Li}, Wei and {Li}, Hong and {Du}, Yuanyuan and {Liu}, Xiaohua and {Meng}, Bin and {Wen}, Xiangyang and {Zhang}, Aimei and {Ma}, Jia and {Li}, Maoshun and {Li}, Gang and {Qi}, Liqiang and {Sun}, Jianchao and {Luo}, Tao and {Liu}, Hongwei and {Liu}, Xiaojing and {Zhang}, Fan and {Luo}, Laidan and {Zhu}, Yuxuan and {Zhao}, Zijian and {Sun}, Liang and {Yang}, Xiongtao and {Wu}, Qiong and {Jiang}, Jiechen and {Shi}, Haoli and {Liu}, Jiangtao and {Xu}, Yanbing and {Yang}, Sheng and {Zhang}, Laiyu and {Han}, Dawei and {Gao}, Na and {Huo}, Jia and {Zhang}, Ziliang and {Wang}, Hao and {Zhao}, Xiaofan and {Wang}, Shuo and {Li}, Zhenjie and {Bao}, Ziyu and {Liu}, Yaoguang and {Wang}, Ke and {Wang}, Na and {Wang}, Bo and {Wang}, Langping and {Wang}, Dianlong and {Ding}, Fei and {Sheng}, Lizhi and {Qiang}, Pengfei and {Yan}, Yongqing and {Liu}, Yongan and {Wu}, Zhenyu and {Liu}, Yichen and {Chen}, Hao and {Zhang}, Yacong and {Liu}, Hongbang and {Altmann}, Alexander and {Bechteler}, Thomas and {Burwitz}, Vadim and {Fiorini}, Carlo and {Friedrich}, Peter and {Meidinger}, Norbert and {Strecker}, Rafael and {Baldini}, Luca and {Bellazzini}, Ronaldo and {Bonino}, Raffaella and {Frass{\`a}}, Andrea and {Latronico}, Luca and {Maldera}, Simone and {Manfreda}, Alberto and {Minuti}, Massimo and {Pesce-Rollins}, Melissa and {Sgr{\`o}}, Carmelo and {Tugliani}, Stefano and {Pareschi}, Giovanni and {Basso}, Stefano and {Sironi}, Giorgia and {Spiga}, Daniele and {Tagliaferri}, Gianpiero and {Tykhonov}, Andrii and {Paltani}, St{\`e}phane and {Bozzo}, Enrico and {Tenzer}, Christoph and {Bayer}, J{\"o}rg and {Tuo}, Youli and {Liu}, Honghui and {Zhang}, Yonghe and {Cai}, Zhiming and {Liu}, Huaqiu and {Chen}, Wen and {Wang}, Chunhong and {He}, Tao and {Chen}, Yehai and {Qiu}, Chengbo and {Zhang}, Ye and {Feng}, Jianchao and {Zhu}, Xiaofei and {Zhou}, Heng and {Zheng}, Shijie and {Song}, Liming and {Wang}, Jinzhou and {Jia}, Shumei and {Jiang}, Zewen and {Li}, Xiaobo and {Zhao}, Haisheng and {Guan}, Ju and {Zhang}, Juan and {Li}, Chengkui and {Huang}, Yue and {Liao}, Jinyuan and {You}, Yuan and {Zhang}, Hongmei and {Wang}, Wenshuai and {Wang}, Shuang and {Ou}, Ge and {Hu}, Hao and {Shi}, Jingyan and {Cui}, Tao and {Jiang}, Xiaowei and {Cheng}, Yaodong and {Li}, Haibo and {Xu}, Yanjun and {Zane}, Silvia and {Bambi}, Cosimo and {Bu}, Qingcui and {Dall'Osso}, Simone and {Rosa}, Alessandra De and {Gou}, Lijun and {Guillot}, Sebastien and {Ji}, Long and {Li}, Ang and {Mao}, Jirong and {Patruno}, Alessandro and {Stratta}, Giulia and {Taverna}, Roberto and {Tsygankov}, Sergey and {Uttley}, Phil and {Watts}, Anna L. and {Wu}, Xuefeng and {Xu}, Renxin and {Yi}, Shuxu and {Zhang}, Guobao and {Zhang}, Liang and {Zhao}, Wen and {Zhou}, Ping},
        title = "{The enhanced X-ray Timing and Polarimetry mission{\textemdash}eXTP for launch in 2030}",
      journal = {Science China Physics, Mechanics, and Astronomy},
         year = 2025,
        month = sep,
       volume = {68},
       number = {11},
          eid = {119502},
        pages = {119502},
          doi = {10.1007/s11433-025-2786-6},
archivePrefix = {arXiv},
       eprint = {2506.08101},
 primaryClass = {astro-ph.HE},
       adsurl = {https://ui.adsabs.harvard.edu/abs/2025SCPMA..6819502Z}
}

@ARTICLE{Ge2025SCPMA,
       author = {{Ge}, Mingyu and {Ji}, Long and {Taverna}, Roberto and {Tsygankov}, Sergey and {Xu}, Yanjun and {Santangelo}, Andrea and {Zane}, Silvia and {Zhang}, Shuang-Nan and {Feng}, Hua and {Chen}, Wei and {Cheng}, Quan and {Hou}, Xian and {Imbrogno}, Matteo and {Israel}, Gian Luca and {Kelly}, Ruth and {Kong}, Ling-Da and {Liu}, Kuan and {Mushtukov}, Alexander and {Poutanen}, Juri and {Suleimanov}, Valery and {Tao}, Lian and {Tong}, Hao and {Turolla}, Roberto and {Wang}, Weihua and {Ye}, Wentao and {Zhao}, Qing-Chang and {Brice}, Nabil and {Geng}, Jinjun and {Lin}, Lin and {Wang}, Wei-Yang and {Xie}, Fei and {Xiong}, Shao-Lin and {Zhang}, Shu and {Fu}, Yucong and {Lai}, Dong and {Li}, Jian and {Li}, Pan-Ping and {Li}, Xiaobo and {Li}, Xinyu and {Liu}, Honghui and {Liu}, Jiren and {Peng}, Jingqiang and {Shui}, Qingcang and {Tuo}, Youli and {Wang}, Hongguang and {Wang}, Wei and {Weng}, Shanshan and {You}, Yuan and {Zheng}, Xiaoping and {Zhou}, Xia},
        title = "{Physics of strong magnetism with eXTP}",
      journal = {Science China Physics, Mechanics, and Astronomy},
         year = 2025,
        month = sep,
       volume = {68},
       number = {11},
          eid = {119505},
        pages = {119505},
          doi = {10.1007/s11433-025-2796-y},
archivePrefix = {arXiv},
       eprint = {2506.08369},
 primaryClass = {astro-ph.HE},
       adsurl = {https://ui.adsabs.harvard.edu/abs/2025SCPMA..6819505G}
}

@ARTICLE{Naghizadeh1993AA,
       author = {{Naghizadeh-Khouei}, J. and {Clarke}, D.},
        title = "{On the statistical behaviour of the position angle of linear polarization}",
      journal = {\aap},
         year = 1993,
        month = jul,
       volume = {274},
        pages = {968},
       adsurl = {https://ui.adsabs.harvard.edu/abs/1993A&A...274..968N}
}

@ARTICLE{Everett2001ApJ,
       author = {{Everett}, J.~E. and {Weisberg}, J.~M.},
        title = "{Emission Beam Geometry of Selected Pulsars Derived from Average Pulse Polarization Data}",
      journal = {\apj},
         year = 2001,
        month = may,
       volume = {553},
       number = {1},
        pages = {341-357},
          doi = {10.1086/320652},
archivePrefix = {arXiv},
       eprint = {astro-ph/0009266},
 primaryClass = {astro-ph},
       adsurl = {https://ui.adsabs.harvard.edu/abs/2001ApJ...553..341E}
}

@ARTICLE{Hogg2010arXiv,
       author = {{Hogg}, David W. and {Bovy}, Jo and {Lang}, Dustin},
        title = "{Data analysis recipes: Fitting a model to data}",
      journal = {arXiv e-prints},
         year = 2010,
        month = aug,
          eid = {arXiv:1008.4686},
        pages = {arXiv:1008.4686},
          doi = {10.48550/arXiv.1008.4686},
archivePrefix = {arXiv},
       eprint = {1008.4686},
 primaryClass = {astro-ph.IM},
       adsurl = {https://ui.adsabs.harvard.edu/abs/2010arXiv1008.4686H}
}

@ARTICLE{Lampton1976ApJ,
       author = {{Lampton}, M. and {Margon}, B. and {Bowyer}, S.},
        title = "{Parameter estimation in X-ray astronomy.}",
      journal = {\apj},
         year = 1976,
        month = aug,
       volume = {208},
        pages = {177-190},
          doi = {10.1086/154592},
       adsurl = {https://ui.adsabs.harvard.edu/abs/1976ApJ...208..177L}
}

@ARTICLE{Taverna2020MNRAS,
       author = {{Taverna}, R. and {Turolla}, R. and {Suleimanov}, V. and {Potekhin}, A.~Y. and {Zane}, S.},
        title = "{X-ray spectra and polarization from magnetar candidates}",
      journal = {\mnras},
         year = 2020,
        month = mar,
       volume = {492},
       number = {4},
        pages = {5057-5074},
          doi = {10.1093/mnras/staa204},
archivePrefix = {arXiv},
       eprint = {2001.07663},
 primaryClass = {astro-ph.HE},
       adsurl = {https://ui.adsabs.harvard.edu/abs/2020MNRAS.492.5057T}
}

@ARTICLE{Fern2007ApJ,
       author = {{Fern{\'a}ndez}, Rodrigo and {Thompson}, Christopher},
        title = "{Resonant Cyclotron Scattering in Three Dimensions and the Quiescent Nonthermal X-ray Emission of Magnetars}",
      journal = {\apj},
         year = 2007,
        month = may,
       volume = {660},
       number = {1},
        pages = {615-640},
          doi = {10.1086/511810},
archivePrefix = {arXiv},
       eprint = {astro-ph/0608281},
 primaryClass = {astro-ph},
       adsurl = {https://ui.adsabs.harvard.edu/abs/2007ApJ...660..615F}
}

@ARTICLE{Johnston2019MNRAS,
       author = {{Johnston}, Simon and {Kramer}, Michael},
        title = "{On the beam properties of radio pulsars with interpulse emission}",
      journal = {\mnras},
         year = 2019,
        month = dec,
       volume = {490},
       number = {4},
        pages = {4565-4574},
          doi = {10.1093/mnras/stz2865},
archivePrefix = {arXiv},
       eprint = {1910.04550},
 primaryClass = {astro-ph.HE},
       adsurl = {https://ui.adsabs.harvard.edu/abs/2019MNRAS.490.4565J}
}

@ARTICLE{Lyne1988MNRAS,
       author = {{Lyne}, A.~G. and {Manchester}, R.~N.},
        title = "{The shape of pulsar radio beams.}",
      journal = {\mnras},
         year = 1988,
        month = oct,
       volume = {234},
        pages = {477-508},
          doi = {10.1093/mnras/234.3.477},
       adsurl = {https://ui.adsabs.harvard.edu/abs/1988MNRAS.234..477L}
}

@ARTICLE{Desvignes2024NatAs,
       author = {{Desvignes}, Gregory and {Weltevrede}, Patrick and {Gao}, Yong and {Jones}, David Ian and {Kramer}, Michael and {Caleb}, Manisha and {Karuppusamy}, Ramesh and {Levin}, Lina and {Liu}, Kuo and {Lyne}, Andrew G. and {Shao}, Lijing and {Stappers}, Ben and {P{\'e}tri}, J{\'e}r{\^o}me},
        title = "{A freely precessing magnetar following an X-ray outburst}",
      journal = {Nature Astronomy},
         year = 2024,
        month = may,
       volume = {8},
        pages = {617-627},
          doi = {10.1038/s41550-024-02226-7},
       adsurl = {https://ui.adsabs.harvard.edu/abs/2024NatAs...8..617D}
}

@ARTICLE{Wang2025ApJ,
       author = {{Wang}, Z. and {Wen}, Z.~G. and {Yuan}, J.~P. and {Wang}, N. and {Han}, W. and {Wang}, H.~G. and {Chen}, J.~L.},
        title = "{Frequency-dependent Emission of the Millisecond Pulsar B1937+21 with the Parkes Ultrawideband Receiver}",
      journal = {\apj},
         year = 2025,
        month = jul,
       volume = {987},
       number = {1},
          eid = {43},
        pages = {43},
          doi = {10.3847/1538-4357/add728},
       adsurl = {https://ui.adsabs.harvard.edu/abs/2025ApJ...987...43W}
}

@ARTICLE{Wen2020ApJ,
       author = {{Wen}, Z.~G. and {Yan}, W.~M. and {Yuan}, J.~P. and {Wang}, H.~G. and {Chen}, J.~L. and {Mijit}, M. and {Yuen}, R. and {Wang}, N. and {Tu}, Z.~Y. and {Dang}, S.~J.},
        title = "{The Mode Switching in Pulsar J1326-6700}",
      journal = {\apj},
         year = 2020,
        month = nov,
       volume = {904},
       number = {1},
          eid = {72},
        pages = {72},
          doi = {10.3847/1538-4357/abbfa3},
archivePrefix = {arXiv},
       eprint = {2011.05526},
 primaryClass = {astro-ph.HE},
       adsurl = {https://ui.adsabs.harvard.edu/abs/2020ApJ...904...72W}
}

@ARTICLE{Wen2026ApJ,
       author = {{Wen}, Zhigang and {Chen}, Jianling and {Yuan}, Jianping and {Wang}, Na and {Yan}, Wenming and {Han}, Wei and {Wang}, Zhen and {Duan}, Xuefeng and {Jing}, Liang and {He}, Pengcheng and {Rusul}, Abdujappar and {Wang}, Hui and {Lyu}, Chengbing},
        title = "{Unveiling the Hidden Radiation: The Persistent Emission of PSR B0823+26 in Its Quiescent State}",
      journal = {\apj},
         year = 2026,
        month = mar,
       volume = {999},
       number = {1},
          eid = {135},
        pages = {135},
          doi = {10.3847/1538-4357/ae3d9c},
       adsurl = {https://ui.adsabs.harvard.edu/abs/2026ApJ...999..135W}
}
\bibliographystyle{aasjournal}

\label{lastpage}
\end{CJK*}
\end{document}